\documentclass[twoside,english,american,sort&compress]{iopart}
\usepackage[T1]{fontenc}
\usepackage[latin9]{inputenc}
\usepackage{geometry}
\usepackage{color}
\usepackage{prettyref}
\usepackage{float}
\usepackage{units}
\usepackage{url}
\usepackage{amstext}
\usepackage{graphicx}
\usepackage[authoryear]{natbib}

\makeatletter

\newcommand{\lyxmathsym}[1]{\ifmmode\begingroup\def\b@ld{bold}
  \text{\ifx\math@version\b@ld\bfseries\fi#1}\endgroup\else#1\fi}

\providecommand{\tabularnewline}{\\}

\usepackage{iopams}
\usepackage{setstack}

\def\newblock{\hskip .11em plus .33em minus .07em}

\newcommand{\eqref}[1]{(\ref{#1})}

\usepackage{soul}
\usepackage{color}
\usepackage{colortbl}
\usepackage{morefloats}
\usepackage{placeins}
\definecolor{lightgray}{gray}{0.8}

\usepackage{ccaption}
\captionnamefont{\small\bf}
\captionstyle{\small}

\usepackage{placeins}

\newrefformat{sec}{\sref{#1}} 
\newrefformat{subsec}{\sref{#1}} 
\newrefformat{fig}{\fref{#1}} 
\newrefformat{tab}{\tref{#1}} 
\newrefformat{eqn}{\eref{#1}} 

\makeatother

\usepackage{babel}
\begin{document}
\foreignlanguage{english}{}
\global\long\def\B{\:\mathrm{B}}%
\foreignlanguage{english}{}
\global\long\def\kB{\:\mathrm{kB}}%
\foreignlanguage{english}{}
\global\long\def\MB{\:\mathrm{MB}}%
\foreignlanguage{english}{}
\global\long\def\GB{\:\mathrm{GB}}%
\foreignlanguage{english}{}
\global\long\def\TB{\:\text{TB}}%
\foreignlanguage{english}{}
\global\long\def\PB{\:\text{PB}}%
\foreignlanguage{english}{}
\global\long\def\MiB{\:\mathrm{MiB}}%
\foreignlanguage{english}{}
\global\long\def\GiB{\:\mathrm{GiB}}%
\foreignlanguage{english}{}
\global\long\def\TiB{\:\text{TiB}}%
\foreignlanguage{english}{}
\global\long\def\PiB{\:\text{PiB}}%
\foreignlanguage{english}{}
\global\long\def\GBps{\:\text{GBps}}%
\foreignlanguage{english}{}
\global\long\def\us{\:\mu\text{sec}}%
\foreignlanguage{english}{}
\global\long\def\bit{\:\mathrm{bit}}%
\foreignlanguage{english}{}
\global\long\def\bits{\:\mathrm{bits}}%
\foreignlanguage{english}{}
\global\long\def\GHz{\:\text{GHz}}%
\foreignlanguage{english}{}
\global\long\def\Hz{\:\text{Hz}}%
\foreignlanguage{english}{}
\global\long\def\rate{\nu}%
\foreignlanguage{english}{}
\global\long\def\indegree{K}%
\foreignlanguage{english}{}
\global\long\def\vps{MT}%
\foreignlanguage{english}{}
\global\long\def\numax{\nu_{\text{max}}}%
\foreignlanguage{english}{}
\global\long\def\ms{\text{\:ms}}%
\foreignlanguage{english}{}
\global\long\def\NE{N_{\text{E}}}%
\foreignlanguage{english}{}
\global\long\def\NI{N_{\text{I}}}%
\foreignlanguage{english}{}
\global\long\def\E{\text{E}}%
\foreignlanguage{english}{}
\global\long\def\I{\text{I}}%
\foreignlanguage{english}{}
\global\long\def\pF{\text{\:pF}}%
\foreignlanguage{english}{}
\global\long\def\mV{\text{\:mV}}%
\foreignlanguage{english}{}
\global\long\def\pA{\text{\:pA}}%
\foreignlanguage{english}{}
\global\long\def\s{\:\mathrm{s}}%
\foreignlanguage{english}{}
\global\long\def\ms{\:\mathrm{ms}}%
\foreignlanguage{english}{}
\global\long\def\FLOPS{\:\mathrm{FLOPS}}%
\foreignlanguage{english}{}
\global\long\def\PFLOPS{\:\text{PFLOPS}}%
\foreignlanguage{english}{}
\global\long\def\TFLOPS{\:\text{TFLOPS}}%
\foreignlanguage{english}{}
\global\long\def\percent{\:\%}%
\foreignlanguage{english}{}
\global\long\def\NSEA{N_{\text{SEA}}}%
\foreignlanguage{english}{}
\global\long\def\NSWP{N_{\text{SWP}}}%

\global\long\def\ensmean#1{\left\langle #1\right\rangle }%

\global\long\def\ensvar#1{\left\langle \delta#1^{2}\right\rangle }%

\global\long\def\cv#1{\text{CV\ensuremath{\left[#1\right]}}}%

\global\long\def\ff#1{\text{FF\ensuremath{\left[#1\right]}}}%

\paper[How to reduce weight resolution]{Prominent characteristics of recurrent neuronal networks are robust
against low synaptic weight resolution}
\author{{\Large{}S Dasbach$^{1}$}\footnotemark\footnotetext{Present address: Forschungszentrum Jülich GmbH, Institut für Energie- und Klimaforschung  Plasmaphysik, Partner of the Trilateral Euregio Cluster (TEC), 52425 Jülich, Germany}{\Large{},
T Tetzlaff$^{1}$, M Diesmann$^{1,2,3}$ and J Senk$^{1}$}}
\address{{\Large{}$^{1}$}{\large{}Institute of Neuroscience and Medicine (INM-6),
Institute for Advanced Simulation (IAS-6), and JARA-Institute Brain
Structure-Function Relationships (INM-10), Jülich Research Centre,
Jülich, Germany}}
\address{{\Large{}$^{2}$}{\large{}Department of Physics, Faculty 1, RWTH Aachen
University, Aachen, Germany}}
\address{{\Large{}$^{3}$}{\large{}Department of Psychiatry, Psychotherapy,
and Psychosomatics, Medical School, RWTH Aachen University, Aachen,
Germany}}
\ead{{\large{}s.dasbach@fz-juelich.de}}
\begin{abstract}
The representation of the natural-density, heterogeneous connectivity
of neuronal network models at relevant spatial scales remains a challenge
for Computational Neuroscience and Neuromorphic Computing. In particular,
the memory demands imposed by the vast number of synapses in brain-scale
network simulations constitutes a major obstacle. Limiting the number
resolution of synaptic weights appears to be a natural strategy to
reduce memory and compute load. In this study, we investigate the
effects of a limited synaptic-weight resolution on the dynamics of
recurrent spiking neuronal networks resembling local cortical circuits,
and develop strategies for minimizing deviations from the dynamics
of networks with high-resolution synaptic weights. We mimic the effect
of a limited synaptic weight resolution by replacing normally distributed
synaptic weights by weights drawn from a discrete distribution, and
compare the resulting statistics characterizing firing rates, spike-train
irregularity, and correlation coefficients with the reference solution.
We show that a naive discretization of synaptic weights generally
leads to a distortion of the spike-train statistics. Only if the weights
are discretized such that the mean and the variance of the total synaptic
input currents are preserved, the firing statistics remains unaffected
for the types of networks considered in this study. For networks with
sufficiently heterogeneous in-degrees, the firing statistics can be
preserved even if all synaptic weights are replaced by the mean of
the weight distribution. We conclude that even for simple networks
with non-plastic neurons and synapses, a discretization of synaptic
weights can lead to substantial deviations in the firing statistics,
unless the discretization is performed with care and guided by a rigorous
validation process. For the network model used in this study, the
synaptic weights can be replaced by low-resolution weights without
affecting its macroscopic dynamical characteristics, thereby saving
substantial amounts of memory.
\end{abstract}

\maketitle

\section{Introduction}

\label{sec:introduction}

Computational neuronal network models constrained by available biological
data constitute a valuable tool for studying brain function. The large
number of neurons in the brain, their dense connectivity, and the
premise that advanced brain functions involve a complex interplay
of different brain regions \citep{Bressler2010_277} pose high computational
demands on model simulations. The human cortex consists of more than
$10^{10}$ neurons \citep{Herculano-Houzel09}, each receiving about
$10^{4}$ connections \citep{Abeles91,Defelipe02}. The requirements
for simulations of networks at this scale by far exceed the limits
of modern workstations. Even on high-performance computing (HPC) systems
that distribute the work load across many compute nodes running designated
simulation software, neuronal networks larger than 10\% of the human
cortex are not accessible to simulation to date \citep{Jordan18_2}.
Studying downscaled networks with reduced neuron and synapse numbers
does not qualify as an alternative to natural-density full-scale networks:
while parameter adjustments can compensate to preserve some characteristics
of the network dynamics such as firing rates, or the sensitivity to
small perturbations \citep{Bachmann20_e1007790}, other features such
as the structure of pairwise correlations in the neuronal activity
cannot be maintained simultaneously \citep{Albada15}.

The complexity of neuronal network models evaluated on conventional
HPC systems is limited by simulation speed and hardware requirements.
Routine simulations of large-scale natural-density networks are still
not a standard. Even with state-of-the-art software and high-performance
machines, simulations of biological processes may take several hundred
times longer than the respective processes in the brain \citep{Jordan18_2}.
Biological processes evolving on long time scales (hours, days, up
to years) such as learning and brain development are therefore impossible
to simulate in reasonable amounts of time. In addition, the power
consumption of large-scale network simulations on HPC systems exceeds
the demands of biological brains by orders of magnitude \citep{VanAlbada18_291}.
In this study, we address another factor obstructing large-scale neuronal
network simulations: the high memory demand \citep{Kunkel2012_5_35,Kunkel14_78}.
In simulations performed with NEST \citep{Gewaltig_07_11204}, a simulation
software optimized for this application area, the required memory
is mainly used for the storage of synapses \citep{Jordan18_2}. While
the network model by \citet{Jordan18_2} involves dynamic synapses
undergoing spike-time dependent plasticity, the problem persists also
for the simplest static synapse models characterized by a constant
weight and transmission delay. Using double-precision floating point
numbers, NEST requires $64$ bit of memory for the weight and $24$
bit for the delay of each synapse \citep{Kunkel14_78}. In the mammalian
neocortex, the number of synapses exceeds the number of neurons by
a factor of $10^{4}$. Hence, even small memory demands for individual
synapses add up to substantial amounts in brain-scale simulations.
Reduced memory consumption leads to faster simulation because the
network model can be represented on fewer compute nodes, thereby reducing
the time required for communication between nodes. Access patterns
of synapses are highly variable due to the random structure of neuronal
networks and their sparse and irregular activity. Therefore, also
on the individual compute nodes a reduced memory consumption helps
as memory access can better be predicted and more of the required
memory fits into the cache.

While these and other limitations may not be overcome using conventional
computers built upon the von-Neumann architecture \citep{Backus78_613,Indiveri15_1379},
the development of novel, brain-inspired hardware architectures promises
a solution. Examples for these so-called neuromorphic hardware systems
with different levels of maturity are SpiNNaker \citep{Furber12_1},
BrainScaleS \citep{Meier15_7409627}, Loihi \citep{Davies18_82},
TrueNorth \citep{Merolla14_668}, and Tianjic \citep{Pei19_106}.
All of these systems are designed after different principles and with
different aims \citep{Furber16_051001}, and they employ different
strategies for handling synaptic weights in an architecture with typically
little available memory. SpiNNaker, for instance, saves the weights
as $16$-bit integer values \citep{Jin09_425} and uses fixed-point
arithmetic for the computations. BrainScaleS, instead, utilizes a
mixed signal approach where the dynamics of individual neurons are
implemented by analog circuits embedded in a silicon wafer and the
weight of each synapse is stored using only $6$-bit \citep{Wunderlich19_260}.
Similarly, GPUs \citep{Knight18_941,Golosio21_627620} and FPGAs
\citep{Gupta15_1737} are used for simulations of neural networks
with reduced numerical precision.

Simulation results obtained with different (neuromorphic) hardware
and software systems are hard to compare \citep{Senk17_243,Gutzen18_90,VanAlbada18_291}.
A number of inherent structural differences (e.g., numerical solvers)
may obscure the role of reduced numerical precision on the network
dynamics. Here, we systematically study the effects of a limited synaptic-weight
resolution in software-based simulations of recurrently connected
spiking neuronal networks. We mimic a limited synaptic-weight resolution
by drawing synaptic weights from a discrete distribution with a predefined
discretization level. All other parameters and dynamical variables
are represented in double precision and all calculations are carried
out using standard double arithmetic in the programming language C++.
An exception are the spike times of the neurons which are bound to
the time grid spanned by the computation time step $h$. This artificially
increases synchronization in the network and introduces a global synchronization
error of first order \citep{Hansel98,Morrison07_47}. The limitation
can be overcome by treating spikes in continuous time. This is more
costly if only a moderate precision is required but leads to shorter
run times of high-precision simulations \citep{Hanuschkin10_113}.
However, in the models considered here the errors are dominated by
other factors \citep{VanAlbada18_291}. Frameworks like NEST may support
both simulation strategies enabling the validation of grid-constrained
results by continuous time simulations with minimal changes to the
executable model description.

In the field of machine learning, a number of previous studies addresses
the effects of low-resolution weights in artificial neural networks
\citep[e.g., ][]{Dundar95_1446,Draghici02_395,Courbariaux_14_arXiv,Gupta15_1737,Muller15_arXiv,Wu16,Guo18_arxiv}.
These studies, however, do not provide any intuitive or theoretical
explanation why a particular weight resolution is sufficient to achieve
a desirable network performance. It is therefore unclear to what extent
the results of these studies generalize to other tasks or networks.
It is particularly difficult to transfer these results to neuroscientific
network models: while in machine learning networks are typically validated
based on the achieved task performance, neuroscience often also focuses
on the idle (``resting state'') or task related network activity.
In this work, we address the origin of potential deviations in the
dynamics of neuronal networks with reduced synaptic-weight resolution
from those obtained with a high-resolution ``reference'' of the
same network, and develop strategies to minimize these deviations.
For some machine learning algorithms such as reservoir computing,
the two views on performance are related as the functional performance
depends on the dynamical characteristics of the underlying neuronal
network. In general, however, task performance is not a predictor
of network dynamics (and vice versa).

We demonstrate our general approach based on variants of the local
cortical microcircuit model by \citet{Potjans14_785}, the ``PD model''.
This model represents the cortical natural-density circuitry underneath
a $\unit[1]{mm^{2}}$ patch of early sensory cortex with almost $80,000$
neurons and $\sim10^{4}$ synapses per neuron, and explains the cell-type
and cortical-layer specific firing statistics observed in nature.
To account for the natural heterogeneity in connection strengths,
the synaptic weights are normally distributed. The PD model may serve
as a building block for brain-size networks, because the fundamental
characteristics of the cortical circuitry at this spatial scale are
similar across different cortical areas and species. In the recent
past, the PD model served as a benchmark for several validation studies
in the rapidly evolving field of Neuromorphic Computing \citep{VanAlbada18_291,Knight18_941,Rhodes19_20190160,Heittmann20_Bernstein,Kurth20_Bernstein,Golosio21_627620}.
With this manuscript, we aim to add the aspect of weight discretization
to the debate.

The manuscript is organized as follows: \Sref{sec:methods} provides
details on the discretization methods, the validation procedure, the
network model, and the network simulations. \Sref{subsec:results-indegree}
exposes the pitfalls of a naive discretization of synaptic weights
and \prettyref{subsec:results-indegree-opt} proposes an optimal discretization
strategy for the given synaptic-weight distribution. For illustration,
sections \ref{subsec:results-indegree} and \ref{subsec:results-indegree-opt}
are based on a variant of the PD model with fixed in-degrees, i.e.,
a network where each neuron within a population receives exactly the
same number of inputs. In \prettyref{subsec:results-total-number},
in contrast, the in-degrees are distributed (as in the original PD
model), allowing for a generalization of the results. \Sref{subsec:theory}
proposes an analytical approach using mean-field theory to substantiate
the simulation results on the role of synaptic-weight and in-degree
distributions. \Sref{subsec:sim-time} investigates the effect of
the simulation duration on the relevance of the employed validation
metrics, and on the validation performance. The final \prettyref{sec:discussion}
summarizes the results and discusses future work towards precise and
efficient neuronal network simulations.

\section{Methods}

\label{sec:methods}

The general approach of this study is to compare simulations of neuronal
networks with differently discretized synaptic weights. To assess
whether the weight discretization influences the network dynamics,
the statistics of the spiking activity in the networks with discretized
weights are compared with the statistics in the reference network
with double precision weights. The following sections describe the
methods used for discretizing the synaptic weights (\prettyref{subsec:discretization})
and for calculating and comparing the network statistics (\prettyref{subsec:Population-statistics}).
\Sref{subsec:network-description} contains specifications of the
neuronal network models employed.

\subsection{Discretization of synaptic weights\label{subsec:discretization}}

Computer number formats determine how many binary digits, i.e.,
bits, of computer memory are occupied by a numerical value and how
these bits are interpreted \citep{Goldberg91_5}. Both the number
of bits, $N_{\text{bits}}$, and their interpretation differ for the
various floating-point and fixed-point formats deployed in software
and hardware. A common format is the IEEE 754 double-precision binary
floating-point format (binary64) which allocates $64$ bits of memory
per value encoding the sign ($1$ bit), the exponent ($11$ bits),
and the significant precision ($52$ bits). In general, the upper
limit of distinguishable values that a format can represent is $2^{N_{\text{bits}}}$.
We here aim to identify a possible lower limit for a bit resolution
required to store the synaptic weights in neuronal network simulations
without compromising the accuracy of the results. The network models
studied in this work assume weights to be sampled from continuous
distributions, yielding values in double precision in the respective
reference implementations.

To mimic a lower bit resolution, we discretize the distributions and
systematically reduce the number of attainable values. On the machine,
the values are still represented in double precision, but the degrees
of discretization considered are by orders of magnitude coarser than
double precision. Our approach is therefore independent of the underlying
number format. For generality and for explicit distinction from the
format-specific $N_{\text{bits}}$, we define the weight resolution
by the number of possible discrete values, $N_{\text{bins}}$, that
a discrete distribution is composed of. In the studied network models,
projections between different pairs of neuronal populations are parameterized
with weights sampled from $N_{\text{distr}}$ distributions, for details
see \prettyref{subsec:network-description}. A weight resolution of
$N_{\text{bins}}$ means that $N_{\text{bins}}$ weight values are
assumed for each of the underlying distributions. The maximum total
number of different weights in a network model with discretized weights
is therefore $N_{\text{bins}}\cdot N_{\text{distr}}$ in addition
to potentially different weights not sampled from a distribution,
e.g., those used to connect external stimulating devices.

After the reference weight values are sampled from the continuous
reference distribution, each one of these sampled weights is subsequently
replaced by one of the $N_{\text{bins}}$ discrete values which are
computed according to a discretization procedure as follows: first
an interval $\left[w_{\text{min}},\,w_{\text{max}}\right]$ is defined.
Then the interval is divided up into $N_{\text{bins}}$ bins of equal
widths such that the left edge of the first bin is $w_{\text{min}}$
and the right edge of the last bin is $w_{\text{max}}$. For each
bin, indexed by $i\in\text{\ensuremath{\left[1,\,...,\,N_{\text{bins}}\right]}}$,
the center value $v_{i}$ is assumed as the discrete value for that
bin:
\begin{equation}
v_{i}=w_{\text{min}}+\left(\frac{1}{2}+i-1\right)\cdot w_{\text{step}}\quad\text{with}\quad w_{\text{step}}=\frac{w_{\text{max}}-w_{\text{min}}}{N_{\text{bins}}}.
\end{equation}
All weights drawn from the continuous reference distribution falling
into a specific bin are replaced by the respective $v_{i}$, meaning
that they are rounded to the nearest discrete value. If a sampled
weight coincides with a bin edge, the larger one of the two possible
$v_{i}$ is chosen. However, the probability that a double precision
weight drawn from a distribution with a continuous probability density
function falls exactly onto the edge of one discrete bin is almost
zero. Weights outside of the interval $\left[w_{\text{min}},\,w_{\text{max}}\right]$
are rounded to the closest discrete values, namely the values of the
first or last bin. The two discretization schemes used in this study
(``naive'' and ``optimized'') differ in their choice of the boundaries
of the interval.

\subsubsection{Naive discretization of normal weight distribution}

Without deeper considerations, it seems reasonable to choose $[w_{\text{min}},\,w_{\text{max}}]$
such that the number of originally drawn weights outside of this interval
is negligible. As we are studying network models in which the underlying
continuous weight distributions are normal distributions with mean
$\overline{w}_{\infty}$ and standard deviation $\Delta w_{\text{\ensuremath{\infty}}}$,
a choice could be:
\begin{equation}
\left[w_{\text{min}},\,w_{\text{max}}\right]=\left[\overline{w}_{\infty}-5\Delta w_{\text{\ensuremath{\infty}}},\,\overline{w}_{\infty}+5\Delta w_{\text{\ensuremath{\infty}}}\right].\label{eq:naive-interval}
\end{equation}

\subsubsection{Optimized discretization of normal weight distribution}

An optimized choice for the boundaries of the interval takes statistical
properties of the discrete weights into account. If the reference
weights are independently generated according to a probability distribution
$p(w)$, the distribution of the discrete weights is a probability
mass function $p^{*}\left(v_{i}\right)=:p_{i}^{*}$ with

\begin{equation}
p_{i}^{*}=\left\{ \begin{array}{lcl}
F\left(w_{\text{min}}+w_{\text{step}}\right)-F\left(-\infty\right) & \text{if} & i=1\\
F\left(w_{\text{min}}+iw_{\text{step}}\right)-F\left(w_{\text{min}}+\left(i-1\right)w_{\text{step}}\right) & \text{if} & i=2,\dots,\left(N_{\text{bins}}-1\right)\\
F\left(\infty\right)-F\left(w_{\text{min}}+\left(N_{\text{bins}}-1\right)w_{\text{step}}\right) & \text{if} & i=N_{\text{bins}}
\end{array}\right.\label{eq:prob-mass-function}
\end{equation}
and $F(w)=\int_{-\infty}^{w}p\left(w'\right)\,\text{d}w'$. The statistical
properties of the discrete weights are calculated as for any other
discrete random variable; mean and standard deviation of the discrete
weights are:

\begin{eqnarray}
\overline{w}_{N_{\text{bins}}} & = & \sum_{i=1}^{N_{\text{bins}}}v_{i}p_{i}^{*}\nonumber \\
\Delta w_{N_{\text{bins}}} & = & \sqrt{\sum_{i=1}^{N_{\text{bins}}}v_{i}^{2}p_{i}^{*}-\overline{w}_{N_{\text{bins}}}^{2}}.\label{eq:mu-sigma}
\end{eqnarray}

\begin{figure}
\centering{}\includegraphics{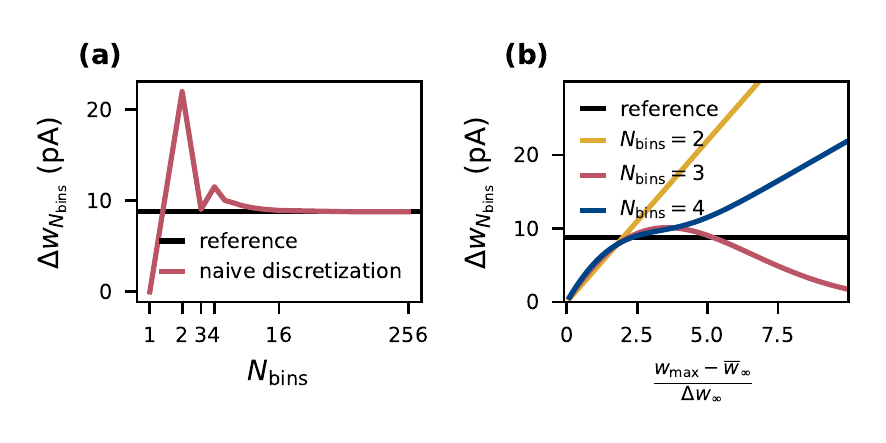}\caption{\textbf{Distortion of synaptic-weight statistics by naive discretization.
}Dependence of the standard deviation $\Delta w_{N_{\text{bins}}}$
of naively discretized synaptic weights (for excitatory connections)
on the number $N_{\text{bins}}$ of bins (a) and on the relative (half-)width
$\left(w_{\text{max}}-\overline{w}_{\infty}\right)/\Delta w_{\text{\ensuremath{\infty}}}$
of the discretization interval (b). The horizontal black line marks
the standard deviation $\Delta w_{\text{\ensuremath{\infty}}}$ of
the corresponding (normal) reference weight distribution.\label{fig:weight-standard-deviation}}
\end{figure}

Due to the symmetry of the underlying normal distribution $p(w)$,
the mean of the discrete distribution $\overline{w}_{N_{\text{bins}}}$
is always equal to the mean of the continuous reference distribution
$\overline{w}_{\infty}$ when placing the bins symmetrically around
$\overline{w}_{\infty}$. On the contrary, the standard deviation
$\Delta w_{N_{\text{bins}}}$ of the discrete weights changes with
the number $N_{\text{bins}}$ of bins (\prettyref{fig:weight-standard-deviation}a).
From \prettyref{eq:prob-mass-function} and \prettyref{eq:mu-sigma}
follows that the standard deviation of the discretized version depends
on the parameters $w_{\text{min}}$, $w_{\text{max}}$ and $N_{\text{bins}}$.
For even numbers of bins, increasing the interval $\left[w_{\text{min}},\,w_{\text{max}}\right]$
causes the standard deviation to diverge to infinity, and for odd
numbers of bins, the standard deviation converges to zero (\prettyref{fig:weight-standard-deviation}b).
Therefore, using even numbers of bins the standard deviation of the
discrete weights $\Delta w_{N_{\text{bins}}}$ matches the reference
standard deviation $\Delta w_{\text{\ensuremath{\infty}}}$ only for
one particular choice of $\left[w_{\text{min}},\,w_{\text{max}}\right]$,
while using odd numbers of bins leads to a second crossing point.
By chance, the naive choice of the interval in \prettyref{eq:naive-interval}
is close to the second intersection for three bins. For high numbers
of bins, the standard deviation is preserved for a wide range of $\left[w_{\text{min}},\,w_{\text{max}}\right]$.

The optimized scheme uses the obtained knowledge of the dependence
on the standard deviation to improve the discretization procedure:
for each number of bins $N_{\text{bins}}$, different interval boundaries
are computed such that the standard deviation is always preserved.

For $N_{\text{bins}}=2$ the optimal choice for $\left[w_{\text{min}},\,w_{\text{max}}\right]$
can be calculated analytically, yielding the interval $\left[\overline{w}_{\infty}-2\Delta w_{\text{\ensuremath{\infty}}},\,\overline{w}_{\infty}+2\Delta w_{\text{\ensuremath{\infty}}}\right]$.
For any higher number of bins, not only $v_{i}$ but also $p_{i}^{*}$
in \prettyref{eq:mu-sigma} depend on the interval, therefore solutions
are found numerically. Here we use Brent's method implemented in \texttt{scipy.optimize.root-scalar}
to find the first intersection. Since the computational effort increases
and yields only negligible gain for higher numbers of bins (\prettyref{fig:weight-standard-deviation}),
the optimization is only performed for $N_{\text{bins}}<2^{16}$
and for higher numbers of bins the fixed interval from \prettyref{eq:naive-interval}
is used. For $N_{\text{bins}}=1$ this optimization is not possible
since the standard deviation is zero by construction.

\subsection{Validation procedure\label{subsec:Population-statistics}}

\subsubsection{Statistics of spiking activity\label{subsec:Statistics-of-spiking}}

We evaluate the effects of discretized synaptic weights on network
dynamics by employing the same statistical spiking-activity characteristics
used in previous studies: distributions of single-neuron firing rates
($\text{FR}$), distributions of coefficients of variation ($\text{CV}$)
of the interspike intervals, and distributions of short-term spike-count
correlation coefficients \citep[CC;][]{Senk17_243,Knight18_941,VanAlbada18_291,Gutzen18_90,Golosio21_627620}.
The time-averaged firing rate 
\[
\text{FR}_{i}=\frac{N_{i}\left(T_{\text{trans}},\,T_{\text{sim}}\right)}{T_{\text{sim}}-T_{\text{trans}}}
\]
of neuron $i$ is defined as the total number $N_{i}\left(T_{\text{trans}},\,T_{\text{sim}}\right)$
of spikes emitted by this particular neuron $i$ during the entire
simulation, i.e., in the time interval $\left[T_{\text{trans}},\,T_{\text{sim}}\right)$,
normalized by the observation duration $(T_{\text{sim}}-T_{\text{trans}})$
\citep{Perkel67a}. The interspike intervals (ISI) are the time intervals
between consecutive spikes of a single neuron. From the ISI distribution,
the coefficient of variation 
\begin{equation}
\text{CV}_{i}=\frac{\sigma_{\text{ISI},i}}{\mu_{\text{ISI},i}}
\end{equation}
 of each neuron $i$ is computed as the ratio between the ISI standard
deviation $\sigma_{\text{ISI},i}$ and its mean $\mu_{\text{ISI},i}$
\citep{Perkel67a}. The CV is a measure of the spike-train irregularity.
In addition to the first-order (single-neuron) measures $\text{FR}_{i}$
and $\text{CV}_{i}$, we quantify the level of synchrony in the network
on short time scales by the Pearson correlation coefficient 
\begin{equation}
\text{CC}_{ij}=\frac{C_{ij}\left(0\right)}{\sqrt{C_{ii}\left(0\right)C_{jj}\left(0\right)}}\label{eq:CC}
\end{equation}
for pairs of neurons $i$ and $j$. Here, 
\begin{equation}
\hspace{-1cm}C_{ij}(\tau)=\left\langle \left(x_{i}\left(t,t+\Delta\right)-\left\langle x_{i}\left(t,t+\Delta\right)\right\rangle _{t}\right)\left(x_{j}\left(t+\tau,t+\Delta+\tau\right)-\left\langle x_{j}\left(t,t+\Delta\right)\right\rangle _{t}\right)\right\rangle _{t}
\end{equation}
 denotes the covariance of the spike counts $x_{i/j}\left(t,\,t+\Delta\right)$
, i.e., the number of spikes in a time interval $[t,\,t+\Delta),$
for a time lag $\tau$ \citep{Perkel67a}. The bin size $\Delta$
for the covariance calculation matches the refractory period of the
neurons in the model networks ($\unit[2]{ms}$). $\text{FR}$, $\text{CV}$
and $\text{CC}$ are calculated using the \texttt{Python} package
\texttt{NetworkUnit} \citep{Gutzen18_90} which relies on the package
\texttt{Elephant} \citep{elephant18}.

\subsubsection{Comparison of distributions\label{subsec:Scores}}

$\text{FR}$ and $\text{CV}$ are calculated for all neurons in each
neuronal population, and the $\text{CC}$ for all pairs of $200$
distinct neurons in each population. The model validation is based
on the distributions of $\text{FR}$, $\text{CV}$ and $\text{CC}$,
respectively, obtained from these ensembles. The distributions are
depicted as histograms with bin sizes $2\cdot\left(\text{IQR }/n^{1/3}\right)$
that are determined using the Freedman-Diaconis rule \citep{Freedman81_453}
based on the inter-quartile range $\text{IQR}$ and the sample size
$n$. For the histograms depicted in figures \ref{fig:indegree-naive},
\ref{fig:indegree-opt}, \ref{fig:total-number-opt}, and \ref{fig:surrogates},
the bin size is calculated for the data obtained from the respective
reference networks with continuous weight distribution, and then used
for all shown distributions for one population. In \prettyref{fig:validation-performance-convergence},
the histogram bin size is obtained from either the longest simulation
($\unit[60]{min}$; a\textendash c), or from the last simulation interval
($\unit[30\lyxmathsym{\textendash}40]{min}$; d\textendash f). While
visual inspection of the histograms yields a qualitative assessment
of the similarity of two distributions $p\left(x\right)$ and $q\left(x\right)$,
the Kolmogorov-Smirnov (KS) score

\begin{equation}
D_{\text{KS}}=\sup\left|P(x)-Q(x)\right|
\end{equation}
provides a quantitative evaluation. The KS score is the maximum vertical
distance between the cumulative distribution functions $P(x)=\int^{x}p\left(y\right)\text{\,d}y$
and $Q(x)=\int^{x}q\left(y\right)\,\text{d}y$ \citep{Gutzen18_90}
and thereby is sensitive to differences in both the shapes and the
positions of the distributions.

The comparison of the distributions of $\text{FR}$, $\text{CV}$
and $\text{CC}$ for a network with continuous weights with those
of a network with discretized weights eliminates other sources of
variability by using the same instantiation of the random network
model. The two networks not only have the same initial conditions,
external inputs, connections between identical pairs of neurons, and
spike-transmission delays: one by one the weights in the discretized
network are the discrete counterparts of the weights in the continuous
network (\prettyref{subsec:discretization}).

\subsection{Description of network models}

\label{subsec:network-description}

The present study uses the model of the cortical microcircuit proposed
by \citet{Potjans14_785}, which mimics the local circuit below $\unit[1]{mm^{2}}$
of cortical surface, as a reference. Tables \ref{tab:network-description}\textendash \ref{tab:network-parameters-cont}
provide a formal description according to \citet{Nordlie-2009_e1000456}.
The PD model organizes the neurons into eight recurrently connected
populations; an excitatory (E) and an inhibitory (I) one in each of
four cortical layers: L2/3E, L2/3I, L4E, L4I, L5E, L5I, L6E, and L6I.
The identical current-based leaky integrate-and-fire dynamics with
exponentially decaying postsynaptic currents describes the neurons
of all populations. Connection probabilities $C_{YX}$ for connections
from population $X$ to population $Y$ are derived from anatomical
and electrophysiological measurements. The weights for the recurrent
synapses are drawn from three different normal distributions ($N_{\text{distr}}=3$):
mean and standard deviation are $\left(\overline{w}_{\infty},\,\Delta w_{\text{\ensuremath{\infty}}}\right)=\unit[\left(87.8,\,8.8\right)]{pA}$
for excitatory and $\left(\overline{w}_{\infty},\,\Delta w_{\text{\ensuremath{\infty}}}\right)=\unit[\left(-351.2,\,35.1\right)]{pA}$
for inhibitory connections. The weights from L4E to L2/3E form an
exception as the values are doubled: $\left(\overline{w}_{\infty},\,\Delta w_{\text{\ensuremath{\infty}}}\right)=\unit[\left(175.6,\,17.6\right)]{pA}$.
To account for Dale's principle \citep{Strata99_349}, negative (positive)
sampled weights of connections that are supposed to be excitatory
(inhibitory) are set to zero. The resulting distributions are therefore
slightly distorted (for the weight distributions used in this study,
this distortion is negligible). Transmission delays are also drawn
from normal distributions with different parameters for excitatory
and inhibitory connections, respectively. Each neuron receives external
input with the statistics of Poisson point process and a constant
weight of $\unit[87.8]{pA}$. The simulations are performed with a
simulation time step of $\unit[0.1]{ms}$ and have a duration $T_{\text{sim}}$
of $15$ biological minutes with exceptions in \prettyref{subsec:sim-time}.
For all simulations, the first second $T_{\text{trans}}=\unit[1]{s}$
is discarded from the analysis. The actual observation time is therefore
$T_{\text{sim}}-T_{\text{trans}}$. For easier readability all times
given in this manuscript always refer to the simulation duration $T_{\text{sim}}$.
The initial membrane potentials of all neurons are randomly drawn
from a population-specific normal distribution to reduce startup transients.

In the original implementation of the model, the total number of synapses
between two populations $S_{YX}$ is derived from an estimate of the
total number of synapses in the volume and exactly $S_{YX}$ synapses
are established. \Sref{subsec:results-total-number} uses this \emph{fixed
total number} connectivity. The \emph{fixed in-degree} network models
in sections \ref{subsec:results-indegree} and \ref{subsec:results-indegree-opt}
determine the in-degrees $K_{YX}$ by dividing the total number of
synapses by the number of neurons in the target population and rounding
up to the next larger integer, see \prettyref{eq:indegrees}. The
rounding ensures that at least one synapse remains for a non-zero
connection probability. \Tref{tab:network-parameters} summarizes
the resulting values of $S_{YX}$ and $K_{YX}$.

\begin{table}
{\small{}}%
\begin{tabular}{|@{\hspace*{1mm}}p{0.2\linewidth}|@{\hspace*{1mm}}p{0.76\linewidth}|}
\hline 
\multicolumn{2}{|l|}{{\small{}\cellcolor{black}}\textbf{\textcolor{white}{\small{}Model
summary}}}\tabularnewline
\hline 
\textbf{\small{}Structure} & {\small{}Multi-layer excitatory-inhibitory ($\text{E}$-$\text{I}$)
network}\tabularnewline
\hline 
\textbf{\small{}Populations} & {\small{}$8$ cortical in $4$ layers ($\text{L2/3}$, $\text{L4}$,
$\text{L5}$, $\text{L6}$)}\tabularnewline
\hline 
\textbf{\small{}Connectivity} & {\small{}Random, independent, population-specific; }\emph{\small{}fixed
in-degree}{\small{} models and }\emph{\small{}fixed total number}{\small{}
models}\tabularnewline
\hline 
\textbf{\small{}Neuron model} & {\small{}Leaky integrate-and-fire (LIF)}\tabularnewline
\hline 
\textbf{\small{}Synapse model} & {\small{}Exponentially shaped postsynaptic currents with normally
distributed static weights}\tabularnewline
\hline 
\textbf{\small{}Input} & {\small{}Independent fixed-rate Poisson spike trains to all neurons
(population-specific in-degree)}\tabularnewline
\hline 
\textbf{\small{}Measurements} & {\small{}Spikes}\tabularnewline
\hline 
\end{tabular}{\small\par}

\caption{Description of the network model following the guidelines of \citet{Nordlie-2009_e1000456}.\label{tab:network-description}}
\end{table}
\begin{table}
{\small{}}%
\begin{tabular}{|@{\hspace*{1mm}}p{0.2\linewidth}|@{\hspace*{1mm}}p{0.76\linewidth}|}
\hline 
\multicolumn{2}{|l|}{{\small{}\cellcolor{black}}\textbf{\textcolor{white}{\small{}Connectivity}}}\tabularnewline
\hline 
\multicolumn{2}{|@{\hspace*{1mm}}p{0.765\linewidth}|}{{\small{}\begin{itemize}\item Connection probabilities $C_{YX}$
from population $X$ to population $Y$ with \newline $\{X,Y\}\in\left\{ \text{L2/3,\,L4,\,L5,\,L6}\right\} \times\left\{ \text{E,I}\right\} $.
Values are given in \citet[table 5]{Potjans14_785}.\item Self-connections
(autapses) are prohibited; multiple connections between neurons (multapses)
are allowed.\end{itemize}}}\tabularnewline
\hline 
\textbf{\emph{\small{}Fixed total number}}\textbf{\small{} models} & {\small{}Total number of synapses \citep[eq. 1]{Potjans14_785}: \begin{equation}S_{YX}=\frac{\log(1-C_{YX})}{\log((N_{Y}N_{X}-1)/(N_{Y}N_{X}))}\label{eq:total_num_synapses}\end{equation}
In- and out-degrees are binomially distributed.}\tabularnewline
\hline 
\textbf{\emph{\small{}Fixed in-degree}}\textbf{\small{}\newline models} & {\small{}In-degree: \begin{equation}K_{YX}=\left\lceil\frac{S_{YX}}{N_{Y}}\right\rceil\label{eq:indegrees}\end{equation}}\tabularnewline
\hline 
\end{tabular}{\small\par}

{\small{}}%
\begin{tabular}{|@{\hspace*{1mm}}p{0.2\linewidth}|@{\hspace*{1mm}}p{0.76\linewidth}|}
\hline 
\multicolumn{2}{|l|}{{\small{}\cellcolor{black}}\textbf{\textcolor{white}{\small{}Neuron
and synapse model}}}\tabularnewline
\hline 
\textbf{\small{}Neuron} & {\small{}Leaky integrate-and-fire neuron (LIF) \begin{itemize}\item
Dynamics of membrane potential $V_{i}\left(t\right)$ for neuron $i$:
\begin{itemize}\item Spike emission at times $t_{s}^{i}$ with $V_{i}\left(t_{s}^{i}\right)\ge V_{\theta}$
\item Subthreshold dynamics with $\tau_{\text{m}}=R_{\text{m}}C_{\text{m}}$:\end{itemize}
\end{itemize}\begin{equation}\tau_{\text{m}}\dot{V}_{i}=-V_{i}+R_{\text{m}}I_{i}\left(t\right)\quad\text{if}\quad\forall s:\,t\notin\left(t_{s}^{i},\,t_{s}^{i}+\tau_{\text{ref}}\right]\label{eq:lif_subthreshold}\end{equation}\begin{itemize}
\item [] \begin{itemize} \item Reset $+$ refractoriness: $V_{i}\left(t\right)=V_{\text{reset}}\quad\text{if}\quad\forall s:\,t\in\left(t_{s}^{i},\,t_{s}^{i}+\tau_{\text{ref}}\right]$
\end{itemize} \item Exact integration with temporal resolution $h$
\citep{Rotter99a} \end{itemize}}\tabularnewline
\hline 
\textbf{\small{}Postsynaptic\newline currents} & {\small{}\begin{itemize}\item Instantaneous onset, exponentially
decaying postsynaptic currents \item Input current of neuron $i$
from presynaptic neuron $j$: \end{itemize} \begin{equation}I_{i}\left(t\right)=\sum_{j}J_{ij}\sum_{s}\text{e}^{-\left(t-t_{s}^{j}-d_{ij}\right)/\tau_{\text{s}}}\Theta\left(t-t_{s}^{j}-d_{ij}\right)\label{eq:lif_psc}\end{equation}}\tabularnewline
\hline 
\textbf{\small{}Synaptic weights\newline (reference\newline distribution)} & {\small{}\begin{itemize} \item Normally distributed (clipped to
preserve sign): \end{itemize} \begin{equation}w_{ij}\sim\mathcal{N}\left\{ \overline{w}_{\infty,YX},\,\Delta w_{\infty,YX}^2\right\}, \,\,\, \overline{w}_{\infty,YX}=g_{YX}\cdot \overline{w}_\infty   \label{eq:weight_distr}\end{equation}}\tabularnewline
\hline 
\textbf{\small{}Spike transmission delays} & {\small{}\begin{itemize} \item Normally distributed (left-clipped
at $h$): \begin{equation}d_{ij}\sim\mathcal{N}\left\{ \overline{d}_X,\,\Delta d_X^2\right\}\label{eq:delay_distr}\end{equation}\end{itemize}}\tabularnewline
\hline 
\textbf{\small{}Initial membrane\newline potentials} & {\small{}\begin{itemize} \item Normally distributed: \begin{equation}V_{ij}\sim\mathcal{N}\left\{ \overline{V}_{0,X},\,\Delta V_{0,X}^2\right\}\label{eq:init_membrane_distr}\end{equation}\end{itemize}}\tabularnewline
\hline 
\end{tabular}{\small\par}

\caption{Description of the network model (continuation of \prettyref{tab:network-description}).\label{tab:network_description_cont}}
\end{table}

\begin{table}
{\small{}}%
\begin{tabular}{|@{\hspace*{1mm}}p{0.05\linewidth}@{\hspace*{1mm}}p{0.073\linewidth}@{\hspace*{1mm}}p{0.07\linewidth}|@{\hspace*{1mm}}p{0.07\linewidth}|@{\hspace*{1mm}}p{0.07\linewidth}|@{\hspace*{1mm}}p{0.07\linewidth}|@{\hspace*{1mm}}p{0.07\linewidth}|@{\hspace*{1mm}}p{0.07\linewidth}|@{\hspace*{1mm}}p{0.07\linewidth}|@{\hspace*{1mm}}p{0.07\linewidth}@{\hspace*{1mm}}p{0.12\linewidth}|}
\hline 
\multicolumn{11}{|l|}{{\small{}\cellcolor{black}}\textbf{\textcolor{white}{\small{}Neuron
and network parameters}}}\tabularnewline
\hline 
\multicolumn{11}{|c|}{{\small{}\cellcolor{lightgray}}\textbf{\small{}Populations and external
in-degree}}\tabularnewline
\hline 
\textbf{\small{}Symbol} & \multicolumn{1}{l|}{} & \multicolumn{1}{l}{\textbf{\small{}Value}} & \multicolumn{1}{l}{} & \multicolumn{1}{l}{} & \multicolumn{1}{l}{} & \multicolumn{1}{l}{} & \multicolumn{1}{l}{} & \multicolumn{1}{l}{} & \multicolumn{1}{l|}{} & \textbf{\small{}Description}\tabularnewline
\hline 
{\small{}$X$} & \multicolumn{1}{l|}{} & {\small{}$\text{L2/3E}$} & {\small{}$\text{L2/3I}$} & {\small{}$\text{L4E}$} & {\small{}$\text{L4I}$} & {\small{}$\text{L5E}$} & {\small{}$\text{L5I}$} & {\small{}$\text{L6E}$} & \multicolumn{1}{l|}{{\small{}$\text{L6I}$}} & {\small{}Population name}\tabularnewline
\hline 
{\small{}$N_{X}$} & \multicolumn{1}{l|}{} & {\small{}$20,683$} & {\small{}$5,834$} & {\small{}$21,915$} & {\small{}$5,479$} & {\small{}$4,850$} & {\small{}$1,065$} & {\small{}$14,395$} & \multicolumn{1}{l|}{{\small{}$2,948$}} & {\small{}Size}\tabularnewline
\hline 
{\small{}$K_{X,\text{ext}}$} & \multicolumn{1}{l|}{} & {\small{}$1,600$} & {\small{}$1,500$} & {\small{}$2,100$} & {\small{}$1,900$} & {\small{}$2,000$} & {\small{}$1,900$} & {\small{}$2,900$} & \multicolumn{1}{l|}{{\small{}$2,100$}} & {\small{}External\newline in-degree}\tabularnewline
\hline 
\multicolumn{11}{|c|}{{\small{}\cellcolor{lightgray}}\textbf{\small{}In-degrees in }\textbf{\emph{\small{}fixed
in-degree}}\textbf{\small{} models}}\tabularnewline
\hline 
{\small{}$K_{YX}$} &  & \multicolumn{8}{c}{{\small{}from~$X$}} & \tabularnewline
 &  & {\small{}$\text{L2/3E}$} & {\small{}$\text{L2/3I}$} & {\small{}$\text{L4E}$} & {\small{}$\text{L4I}$} & {\small{}$\text{L5E}$} & {\small{}$\text{L5I}$} & {\small{}$\text{L6E}$} & {\small{}$\text{L6I}$} & \tabularnewline
\cline{3-10} \cline{4-10} \cline{5-10} \cline{6-10} \cline{7-10} \cline{8-10} \cline{9-10} \cline{10-10} 
 & {\small{}$\text{L2/3E}$} & {\small{}2,200} & {\small{}1,080} & {\small{}980} & {\small{}468} & {\small{}160} & {\small{}0} & {\small{}110} & {\small{}0} & \tabularnewline
\cline{3-10} \cline{4-10} \cline{5-10} \cline{6-10} \cline{7-10} \cline{8-10} \cline{9-10} \cline{10-10} 
 & {\small{}$\text{L2/3I}$} & {\small{}2,991} & {\small{}861} & {\small{}704} & {\small{}290} & {\small{}381} & {\small{}0} & {\small{}61} & {\small{}0} & \tabularnewline
\cline{3-10} \cline{4-10} \cline{5-10} \cline{6-10} \cline{7-10} \cline{8-10} \cline{9-10} \cline{10-10} 
 & {\small{}$\text{L4E}$} & {\small{}160} & {\small{}35} & {\small{}1,118} & {\small{}795} & {\small{}33} & {\small{}1} & {\small{}668} & {\small{}0} & \tabularnewline
\cline{3-10} \cline{4-10} \cline{5-10} \cline{6-10} \cline{7-10} \cline{8-10} \cline{9-10} \cline{10-10} 
{\small{}to~$Y$} & {\small{}$\text{L4I}$} & {\small{}1,481} & {\small{}17} & {\small{}1,814} & {\small{}954} & {\small{}17} & {\small{}0} & {\small{}1,609} & {\small{}0} & \tabularnewline
\cline{3-10} \cline{4-10} \cline{5-10} \cline{6-10} \cline{7-10} \cline{8-10} \cline{9-10} \cline{10-10} 
 & {\small{}$\text{L5E}$} & {\small{}2,189} & {\small{}375} & {\small{}1,136} & {\small{}32} & {\small{}421} & {\small{}497} & {\small{}297} & {\small{}0} & \tabularnewline
\cline{3-10} \cline{4-10} \cline{5-10} \cline{6-10} \cline{7-10} \cline{8-10} \cline{9-10} \cline{10-10} 
 & {\small{}$\text{L5I}$} & {\small{}1,166} & {\small{}160} & {\small{}571} & {\small{}13} & {\small{}301} & {\small{}405} & {\small{}125} & {\small{}0} & \tabularnewline
\cline{3-10} \cline{4-10} \cline{5-10} \cline{6-10} \cline{7-10} \cline{8-10} \cline{9-10} \cline{10-10} 
 & {\small{}$\text{L6E}$} & {\small{}326} & {\small{}39} & {\small{}468} & {\small{}92} & {\small{}286} & {\small{}22} & {\small{}582} & {\small{}753} & \tabularnewline
\cline{3-10} \cline{4-10} \cline{5-10} \cline{6-10} \cline{7-10} \cline{8-10} \cline{9-10} \cline{10-10} 
 & {\small{}$\text{L6I}$} & {\small{}767} & {\small{}6} & {\small{}75} & {\small{}3} & {\small{}137} & {\small{}9} & {\small{}980} & {\small{}460} & \tabularnewline
\hline 
\multicolumn{11}{|c|}{{\small{}\cellcolor{lightgray}}\textbf{\small{}Total number of synapses
in }\textbf{\emph{\small{}fixed total number}}\textbf{\small{} models}}\tabularnewline
\hline 
{\small{}$S_{YX}$} &  & \multicolumn{8}{c}{{\small{}from~$X$}} & \tabularnewline
 &  & {\small{}$\text{L2/3E}$} & {\small{}$\text{L2/3I}$} & {\small{}$\text{L4E}$} & {\small{}$\text{L4I}$} & {\small{}$\text{L5E}$} & {\small{}$\text{L5I}$} & {\small{}$\text{L6E}$} & {\small{}$\text{L6I}$} & \tabularnewline
\cline{3-10} \cline{4-10} \cline{5-10} \cline{6-10} \cline{7-10} \cline{8-10} \cline{9-10} \cline{10-10} 
 & {\small{}$\text{L2/3E}$} & {\tiny{}45,499,804} & {\tiny{}22,323,576} & {\tiny{}20,253,647} & {\tiny{}9,670,918} & {\tiny{}3,293,577} & {\tiny{}0} & {\tiny{}2,271,403} & {\tiny{}0} & \tabularnewline
\cline{3-10} \cline{4-10} \cline{5-10} \cline{6-10} \cline{7-10} \cline{8-10} \cline{9-10} \cline{10-10} 
 & {\small{}$\text{L2/3I}$} & {\tiny{}17,443,694} & {\tiny{}5,018,762} & {\tiny{}4,105,338} & {\tiny{}1,690,073} & {\tiny{}2,221,212} & {\tiny{}0} & {\tiny{}353,460} & {\tiny{}0} & \tabularnewline
\cline{3-10} \cline{4-10} \cline{5-10} \cline{6-10} \cline{7-10} \cline{8-10} \cline{9-10} \cline{10-10} 
 & {\small{}$\text{L4E}$} & {\tiny{}3,503,669} & {\tiny{}756,561} & {\tiny{}24,482,849} & {\tiny{}17,413,575} & {\tiny{}714,524} & {\tiny{}7,002} & {\tiny{}14,624,431} & {\tiny{}0} & \tabularnewline
\cline{3-10} \cline{4-10} \cline{5-10} \cline{6-10} \cline{7-10} \cline{8-10} \cline{9-10} \cline{10-10} 
{\small{}to~$Y$} & {\small{}$\text{L4I}$} & {\tiny{}8,114,253} & {\tiny{}92,831} & {\tiny{}9,933,537} & {\tiny{}5,223,271} & {\tiny{}87,836} & {\tiny{}0} & {\tiny{}8,810,905} & {\tiny{}0} & \tabularnewline
\cline{3-10} \cline{4-10} \cline{5-10} \cline{6-10} \cline{7-10} \cline{8-10} \cline{9-10} \cline{10-10} 
 & {\small{}$\text{L5E}$} & {\tiny{}10,613,575} & {\tiny{}1,817,058} & {\tiny{}5,507,804} & {\tiny{}151,900} & {\tiny{}2,040,738} & {\tiny{}2,407,889} & {\tiny{}1,438,969} & {\tiny{}0} & \tabularnewline
\cline{3-10} \cline{4-10} \cline{5-10} \cline{6-10} \cline{7-10} \cline{8-10} \cline{9-10} \cline{10-10} 
 & {\small{}$\text{L5I}$} & {\tiny{}1,241,436} & {\tiny{}169,424} & {\tiny{}607,666} & {\tiny{}12,851} & {\tiny{}319,601} & {\tiny{}430,443} & {\tiny{}132,414} & {\tiny{}0} & \tabularnewline
\cline{3-10} \cline{4-10} \cline{5-10} \cline{6-10} \cline{7-10} \cline{8-10} \cline{9-10} \cline{10-10} 
 & {\small{}$\text{L6E}$} & {\tiny{}4,681,225} & {\tiny{}556,108} & {\tiny{}6,727,569} & {\tiny{}1,320,233} & {\tiny{}4,112,224} & {\tiny{}305,028} & {\tiny{}837,2649} & {\tiny{}10,827,677} & \tabularnewline
\cline{3-10} \cline{4-10} \cline{5-10} \cline{6-10} \cline{7-10} \cline{8-10} \cline{9-10} \cline{10-10} 
 & {\small{}$\text{L6I}$} & {\tiny{}2,260,836} & {\tiny{}17,207} & {\tiny{}220,032} & {\tiny{}8,078} & {\tiny{}401,637} & {\tiny{}25,217} & {\tiny{}2,888,426} & {\tiny{}1,354,319} & \tabularnewline
\hline 
\end{tabular}{\small\par}

{\small{}}%
\begin{tabular}{|@{\hspace*{1mm}}p{0.1\linewidth}@{}|@{\hspace*{1mm}}p{0.15\linewidth}|@{\hspace*{1mm}}p{0.7\linewidth}|}
\hline 
\multicolumn{3}{|c|}{{\small{}\cellcolor{lightgray}}\textbf{\small{}Connection parameters
and external input}}\tabularnewline
\hline 
\textbf{\small{}Symbol} & \textbf{\small{}Value} & \textbf{\small{}Description}\tabularnewline
\hline 
{\small{}$\overline{w}_{\infty}$} & {\small{}$\unit[87.81]{pA}$} & {\small{}Reference synaptic strength. All synapse weights are measured
in units of $\overline{w}_{\infty}$.}\tabularnewline
{\small{}$g_{YX}$} &  & {\small{}Relative synaptic strengths:}\tabularnewline
 & {\small{}$1$} & {\small{}$X\in\left\{ \text{\text{L2/3E,\,L4E,\,L5E,\,L6E}}\right\} $}\tabularnewline
 & {\small{}$-4$} & {\small{}$\ensuremath{X\in\{\text{L2/3I,\,L4I,\,L5I,\,L6I}\}}$, except
for:}\tabularnewline
 & {\small{}$2$} & {\small{}$\left(X,Y\right)=\left(\text{L4E,\,L2/3E}\right)$}\tabularnewline
{\small{}$\Delta w_{\infty,YX}$} & {\small{}$0.1\cdot g_{YX}\cdot\overline{w}_{\infty}$} & {\small{}Standard deviation of weight distribution}\tabularnewline
{\small{}$\overline{d}_{\text{E}}$} & {\small{}$\unit[1.5]{ms}$} & {\small{}Mean excitatory delay}\tabularnewline
{\small{}$\overline{d}_{\text{\ensuremath{I}}}$} & {\small{}$\unit[0.75]{ms}$} & {\small{}Mean inhibitory delay}\tabularnewline
{\small{}$\Delta d_{X}$} & {\small{}$0.5\cdot\overline{d}_{X}$} & {\small{}Standard deviation of delay distribution}\tabularnewline
{\small{}$\nu_{\text{ext}}$} & {\small{}$\unit[8]{s^{-1}}$} & {\small{}Rate of external input with Poisson }interspike{\small{}
interval statistics}\tabularnewline
\hline 
\multicolumn{3}{|c|}{{\small{}\cellcolor{lightgray}}\textbf{\small{}LIF neuron model}}\tabularnewline
\hline 
\textbf{\small{}Symbol} & \textbf{\small{}Value} & \textbf{\small{}Description}\tabularnewline
\hline 
{\small{}$C_{\text{m}}$} & {\small{}$\unit[250]{pF}$} & {\small{}Membrane capacitance}\tabularnewline
{\small{}$\tau_{\text{m}}$} & {\small{}$\unit[10]{ms}$} & {\small{}Membrane time constant}\tabularnewline
{\small{}$E_{\text{L}}$} & {\small{}$\unit[-65]{mV}$} & {\small{}Resistive leak reversal potential}\tabularnewline
{\small{}$V_{\theta}$} & {\small{}$\unit[-50]{mV}$} & {\small{}Spike detection threshold}\tabularnewline
{\small{}$V_{\text{reset}}$} & {\small{}$\unit[-65]{mV}$} & {\small{}Spike reset potential}\tabularnewline
{\small{}$\tau_{\text{ref}}$} & {\small{}$\unit[2]{ms}$} & {\small{}Absolute refractory period after spikes}\tabularnewline
{\small{}$\tau_{\text{s}}$} & {\small{}$\unit[0.5]{ms}$} & {\small{}Postsynaptic current time constant}\tabularnewline
\hline 
\end{tabular}{\small\par}

\caption{Neuron, network, and simulation parameters. \label{tab:network-parameters}}
\end{table}

\begin{table}
{\small{}}%
\begin{tabular}{|@{\hspace*{1mm}}p{0.05\linewidth}@{\hspace*{1mm}}p{0.05\linewidth}|@{\hspace*{1mm}}p{0.07\linewidth}|@{\hspace*{1mm}}p{0.07\linewidth}|@{\hspace*{1mm}}p{0.07\linewidth}|@{\hspace*{1mm}}p{0.07\linewidth}|@{\hspace*{1mm}}p{0.07\linewidth}|@{\hspace*{1mm}}p{0.07\linewidth}|@{\hspace*{1mm}}p{0.07\linewidth}|@{\hspace*{1mm}}p{0.07\linewidth}|@{\hspace*{1mm}}p{0.12\linewidth}|}
\hline 
\multicolumn{11}{|l|}{{\small{}\cellcolor{black}}\textbf{\textcolor{white}{\small{}Neuron
and network parameters (cont.)}}}\tabularnewline
\hline 
\multicolumn{11}{|c|}{{\small{}\cellcolor{lightgray}}\textbf{\small{}Initial membrane potentials}}\tabularnewline
\hline 
\textbf{\small{}Symbol} &  & \multicolumn{1}{l}{\textbf{\small{}Value}} & \multicolumn{1}{l}{} & \multicolumn{1}{l}{} & \multicolumn{1}{l}{} & \multicolumn{1}{l}{} & \multicolumn{1}{l}{} & \multicolumn{1}{l}{} &  & \textbf{\small{}Description}\tabularnewline
\hline 
{\small{}$X$} &  & {\small{}$\text{L2/3E}$} & {\small{}$\text{L2/3I}$} & {\small{}$\text{L4E}$} & {\small{}$\text{L4I}$} & {\small{}$\text{L5E}$} & {\small{}$\text{L5I}$} & {\small{}$\text{L6E}$} & {\small{}$\text{L6I}$} & {\small{}Population name}\tabularnewline
\hline 
{\small{}$\overline{V}_{0,X}$} &  & {\small{}$-68.28$} & {\small{}$-63.16$} & {\small{}$-63.33$} & {\small{}$-63.45$} & {\small{}$-63.11$} & {\small{}$-61.66$} & {\small{}$-66.72$} & {\small{}$-61.43$} & {\small{}Mean in mV}\tabularnewline
\hline 
{\small{}$\Delta V_{0,X}$} &  & {\small{}$5.36$} & {\small{}$4.57$} & {\small{}$4.74$} & {\small{}$4.94$} & {\small{}$4.94$} & {\small{}$4.55$} & {\small{}$5.46$} & {\small{}$4.48$} & {\small{}Standard\newline deviation \newline in mV}\tabularnewline
\hline 
\end{tabular}{\small\par}

{\small{}}%
\begin{tabular}{|@{\hspace*{1mm}}p{0.1\linewidth}@{}|@{\hspace*{1mm}}p{0.15\linewidth}|@{\hspace*{1mm}}p{0.7\linewidth}|}
\hline 
\multicolumn{3}{|l|}{{\small{}\cellcolor{black}}\textbf{\textcolor{white}{\small{}Simulation
parameters}}}\tabularnewline
\hline 
\textbf{\small{}Symbol} & \textbf{\small{}Value} & \textbf{\small{}Description}\tabularnewline
\hline 
{\small{}$T_{\text{sim}}$} & {\small{}$\unit[15]{min}$} & {\small{}Simulation duration}\tabularnewline
{\small{}$h$} & {\small{}$\unit[0.1]{ms}$} & {\small{}Temporal resolution}\tabularnewline
{\small{}$T_{\text{trans}}$} & {\small{}$\unit[1]{s}$} & {\small{}Startup transient}\tabularnewline
\hline 
\end{tabular}{\small\par}

\caption{Neuron, network, and simulation parameters (continuation of \prettyref{tab:network-parameters}).\label{tab:network-parameters-cont}}
\end{table}

\subsection{Software environment and simulation architecture}

The simulations in this study are performed on the JURECA supercomputer
at the J\"{u}lich Research Centre, Germany. JURECA consists of 1872
compute nodes, each with two Intel Xeon E5-2680 v3 Haswell CPUs running
at 2.5$\GHz$. The processors have 12 cores and support 2 hardware
threads per core. Each compute node has at least 128$\GB$ of memory
available. The compute nodes are connected via Mellanox EDR InfiniBand.

All neural network simulations in this study are performed using the
\texttt{NEST} simulation software \citep{Gewaltig_07_11204}. NEST
uses double precision floating point numbers for the network parameters
and the calculations. The simulation kernel is written in C++ but
the simulations are defined via the Python interface \texttt{PyNEST}
\citep{Eppler09_12}. The simulations of the cortical microcircuit
are performed with \texttt{NEST} compiled from the master branch (commit
8adec3c, \url{https://github.com/nest/nest-simulator}). The compilations
are performed with the GNU Compiler Collection (GCC). ParaStationMPI
library is used for MPI support. Each simulation runs on a single
compute node with 1 MPI process and 24 OpenMP threads.

All analyses are carried out with \texttt{Python} 3.6.8 and the following
packages: \texttt{NumPy} (version 1.15.2), \texttt{SciPy} (version
1.2.1), \texttt{Matplotlib} (version 3.0.3), \texttt{Elephant} (version
0.5.0, \url{https://python-elephant.org}), and \texttt{NetworkUnit}
(version 0.1.0, \url{https://github.com/INM-6/NetworkUnit}).

The source code to reproduce all figures of this manuscript is publicly
available at \url{https://doi.org/10.5281/zenodo.4696168}.

\FloatBarrier

\section{Results}

\label{sec:results}

In this study, the evaluation of the role of the synaptic weight
resolution is based on the model of a local cortical microcircuit
derived by \citet{Potjans14_785}. The model comprises four cortical
layers (L2/3, L4, L5, and L6), each containing an excitatory (E) and
an inhibitory (I) neuron population. An $8\times8$ matrix of cell-type
and layer specific connection probabilities provides the basis of
the connectivity between neurons (table 5 in \citealp{Potjans14_785}).
Based on this matrix, the present manuscript considers two different
probabilistic algorithms to determine which individual neurons in
any pair of populations are being connected. First, \prettyref{subsec:results-indegree}
uses a \emph{fixed in-degree} rule which requires for each neuron
of a target population the same number of incoming connections from
a source population. Second, in \prettyref{subsec:results-total-number}
the total number of synapses between two populations is calculated
and synapses are established successively until this number is reached.
We refer to this latter procedure, which was also employed in the
original implementation by \citet{Potjans14_785}, as the \emph{fixed
total number} rule. In both algorithms, synapses are drawn randomly;
the exact connectivity realization is hence dependent on the specific
sequence of random numbers required for the sampling process, i.e.,
the choice and the seed of the employed pseudo-random number generators.

In the PD model, a spike of a presynaptic neuron elicits, after a
transmission delay, a jump in the synaptic currents of its postsynaptic
targets which decays exponentially with time. In the original implementation,
the synaptic weights, the amplitudes of these jumps, are drawn from
normal distributions when connections are established, and they remain
constant for the course of the following state-propagation phase.
All excitatory weights are sampled from a normal distribution with
the same (positive) mean and the same standard deviation, except for
connections from L4E to L2/3E where the mean and standard deviation
are doubled. All inhibitory weights are sampled with a different (negative)
mean and a different standard deviation.

This study compares the activity statistics obtained from simulations
of a reference model with continuous weight distributions with those
where the synaptic weights are drawn from the same continuous distributions
and subsequently discretized. We refer to an ``$N_{\text{bins}}$
discretization'' as the case where the sampled weights are replaced
by a finite set of $N_{\text{bins}}\in\mathcal{\mathbb{N}}^{+}$ discrete
values for each of the three weight distributions. As validation measures,
we use the time-averaged single-neuron firing rates ($\text{FR}$),
the coefficients of variation ($\text{CV}$) of the interspike intervals
as a spike-train irregularity measure, and the short-term spike-train
correlation coefficients ($\text{CC}$) as a synchrony measure. We
quantify the discretization error, i.e., the deviation between the
discretized and the reference model, by the Kolmogorov-Smirnov (KS)
score $D_{\text{KS}}$ computed from the empirical distributions of
these statistical measures across neurons. To evaluate the significance
of the discretization error, we recognize that the model is defined
in a probabilistic manner: valid predictions of this model are those
features that are exhibited by the ensemble of model realizations.
Features that are specific to a single realization are meaningless.
Therefore, deviations between realizations of a discretized and the
reference model are significant only if they exceed those between
different realizations of the reference model. In other words, if
the observed KS score between the discretized and the reference model
falls into the distribution of KS scores obtained from an ensemble
of pairs of reference realizations, the weight discretization does
not lead to significant errors with respect to the considered validation
measure.

\subsection{Naive discretization distorts statistics of spiking activity}

\label{subsec:results-indegree}

\begin{figure}[p]
\begin{centering}
\includegraphics[width=1\textwidth]{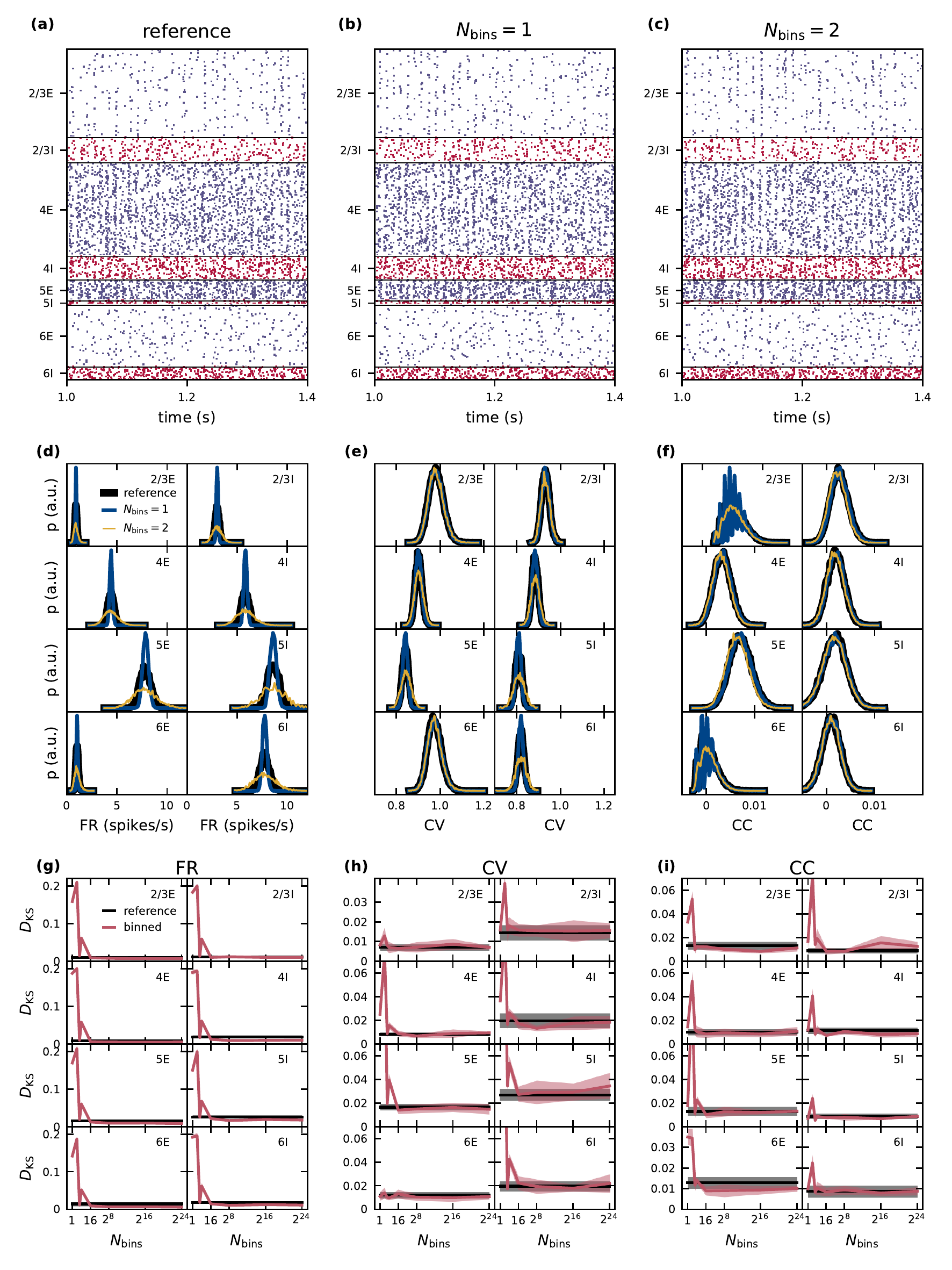}
\par\end{centering}
\caption{\textbf{Effect of naive weight discretization on the spike-train statistics
in networks with fixed in-degrees.} Caption continued on next page.\label{fig:indegree-naive}}
\end{figure}

\begin{figure}[t] \contcaption{(a)\textendash (c): Spiking activity
(dots mark time and sender of each spike) of $5\%$ of all excitatory
(blue) and inhibitory (red) neurons of the eight neuronal populations
(vertically arranged) of the PD model with fixed in-degrees. Spike
times from simulations of the reference network (a) and of networks
with naively discretized $1$-bin (b) and $2$-bin weights (c). (d)\textendash (f):
Population-specific distributions of single-neuron firing rates $\text{FR}$
(d), coefficients of variation $\text{CV}$ of the interspike intervals
(e), and spike-train correlation coefficients $\text{CC}$ (f) from
simulations of the reference network (black), as well as networks
with $1$-bin (blue) and $2$-bin weights (yellow). (g)\textendash (i):
Mean (solid curves) and standard deviation (shaded areas) of the Kolmogorov-Smirnov
scores $D_{\text{KS}}$ obtained from distributions in panels (d)\textendash (f)
across five different network realizations. Red: Comparison of simulation
results with discretized ($N_{\text{bins}}=1,\ldots,2^{24}$ ) and
reference weights (with identical random-number generator seed). Black:
Comparison of different random realizations of the reference network.}\end{figure}

The connectivity of the PD model exhibits different sources of heterogeneity:
connections between pairs of neurons result from a random process
and distributions govern the creation of their weights and delays.
A number of previous studies have shown how such heterogeneities influence
neuronal network dynamics \citep{Tsodyks1993_1280,Golomb93_4810,VanVreeswijk98_1321,neltner2000_1607,Denker04,Roxin11_16217,Roxin11_8,Pfeil16_021023}.
In particular, distributed in-degrees, as implemented with the \emph{fixed
total number} rule in the original version of the model by \citet{Potjans14_785},
can obscure effects of altered weight distributions which are the
primary subject of this study. To isolate the role of the weight distribution,
we therefore start by investigating a \textit{fixed in-degree} version
of the PD model. To assess how a discretization of the weights affects
the spiking activity in the network, we begin with a simple ``naive''
discretization scheme: an arbitrary interval is defined around the
mean value of the underlying normal distribution (here: $\pm5$ standard
deviations) and discretized into a desired number of bins. Each weight
sampled from the continuous distribution is replaced by the nearest
bin value, respectively (for details, see \prettyref{subsec:discretization}).

We use similar measures and procedures as previous studies (e.g.,
\citealp{VanAlbada18_291,Knight18_941}) to compare the activity on
a statistical level, but with the major difference that here the network
is simulated longer, in fact $15$ minutes of biological time (see
\prettyref{subsec:sim-time}). The raster plots in \prettyref{fig:indegree-naive}(a\textendash c)
show qualitatively similar asynchronous irregular spiking activity
in all neuronal populations. The individual spike times, however,
are different in the networks with synaptic weights using the reference
implementation with double precision in panel (a) and in the networks
with $1$- and $2$-bin weights in panels (b) and (c), respectively.
The dynamics of recurrent neuronal networks similar to the PD model
is often chaotic \citep{Sompolinsky88_259,VanVreeswijk98_1321,Monteforte10_268104}.
Even tiny perturbations (such as modifications in synaptic weights)
can therefore cause large deviations in the microscopic dynamics.
Macroscopic characteristics such as distributions of firing rates,
spike-train regularity and synchrony measures, however, should not
be affected. Preserving the spiking statistics upon weight discretization
is therefore an aim of this study.

The distributions of time-averaged firing rates obtained with $1$-bin
weights have a similar mean as the reference distribution, but are
more narrow in all populations (\prettyref{fig:indegree-naive}d).
In homogeneous networks with non-distributed $1$-bin weights, analytical
studies predict that all neurons inside one population have the same
firing rate \citep{Brunel00_183,Helias14}, in contrast to the reference
network with distributed weights and an expected rate distribution
of finite width. The remaining finite width of the rate distribution
obtained from network simulations with $1$-bin weights is a finite-size
effect and decreases further for larger networks and longer simulation
times. For $2$-bin weights generated by this naive discretization
scheme, the rate distributions are broader than in the reference network
(\prettyref{fig:indegree-naive}d). For several neuronal populations,
such as L2/3E or L2/3I, the distributions of the coefficients of variation
of the interspike intervals obtained from networks with discrete weights
are similar to those of the reference network (\prettyref{fig:indegree-naive}e).
In other populations, such as L6I, the $\text{CV}$ distributions
are narrower for $1$-bin weights and broader for $2$-bin weights,
while the mean is preserved. The distributions of correlation coefficients
in the discretized implementations are similar to the reference version
for most populations (\prettyref{fig:indegree-naive}f). Only in L2/3E
and L6E we see an oscillatory pattern for one bin in the region of
small correlations. The same oscillatory pattern is also present in
the $\text{CC}$ distribution of the reference network but less pronounced
(not visible here).

To quantify the differences in the resulting distributions we use
the Kolmogorov-Smirnov ($\text{KS}$) score. In each case we compare
the distributions of FR, CV and CC obtained from simulations of networks
with binned weights to the reference distributions. To assess the
significance of non-zero $\text{KS}$ scores, we repeat the comparison
analysis for pairs of (random) realizations of the reference network
(i.e., different realizations of the connectivity, spike-transmission
delays, external inputs, and initial conditions). As simulation results
should not qualitatively depend on the specific realization of the
probabilistically defined model, all deviations ($\text{KS}$ scores)
which are of the same size as or smaller than this baseline are insignificant.
For all three activity statistics ($\text{FR}$, $\text{CV}$, $\text{CC}$)
the deviations are largest for one and two bins (\prettyref{fig:indegree-naive}g\textendash i).
For around $16$ bins the deviations in all three activity statistics
converge towards a non-zero $\text{KS}$ score, and do not decrease
further with any higher number of bins. This residual deviation is
the minimal possible deviation for this simulation time. In the \emph{fixed
in-degree} network using the naive discretization scheme, these deviations
are smaller than the baseline obtained using different network realizations
from $16$ bins onward. For the kind of network simulation studied
here and the specific choice of the binning, $16$ bins are therefore
sufficient to achieve activity statistics with a satisfactory precision.
For lower numbers of bins, a pattern appears in all populations and
for all three statistical measures: the deviations from the reference
network do not decrease monotonously with increasing number of bins,
but increase from one to two bins, decrease from two to three, and
increase again from three to four bins (\prettyref{fig:indegree-naive}g\textendash i).
These differences are highly significant as in several neuronal populations
three bins achieve a score value better than the reference obtained
using different seeds while four bins do not. The weight discretization
procedure (\prettyref{subsec:discretization}) reveals a hint on the
origin of this behavior. The naive discretization scheme changes the
standard deviation of the weight distributions depending on the number
of bins. Three bins achieve a good result by a mere coincidence, because
due to the choice of the binning interval, the standard deviation
of the discrete weights is close to the standard deviation of the
reference distributions (see \prettyref{fig:weight-standard-deviation}).
Comparing the KS scores in \prettyref{fig:indegree-naive}(g) with
the discrepancies between the standard deviations in \prettyref{fig:weight-standard-deviation}(a)
exhibits the same pattern in both measures.

\subsection{Optimized discretization preserves statistics of spiking activity}

\label{subsec:results-indegree-opt}

\begin{figure}
\begin{centering}
\includegraphics[width=1\textwidth]{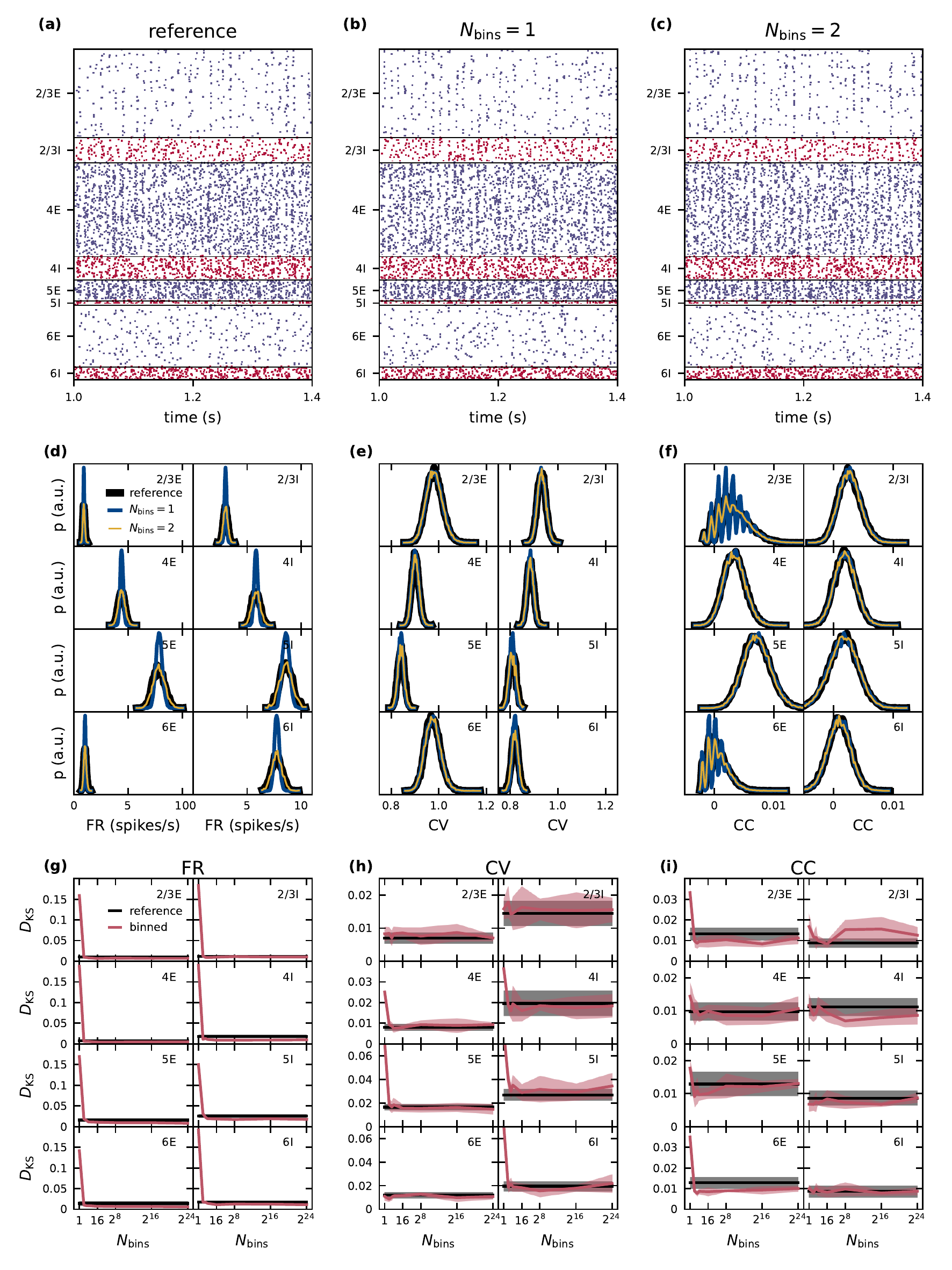}
\par\end{centering}
\caption{\textbf{Effect of optimized weight discretization on the spike-train
statistics in networks with fixed in-degrees. }Same display as in
\prettyref{fig:indegree-naive}. \label{fig:indegree-opt}}
\end{figure}

Suspecting that a discrepancy between the standard deviations of the
weight distributions in the reference and the binned network results
in deviant activity statistics, we derive a discretization method
that preserves the standard deviation of the reference weights for
any number of bins. This method adapts the width of the interval in
which the discrete bins are evenly placed, depending on the number
of bins and the reference weight distribution (\prettyref{subsec:discretization}).
If the discrepancy in the standard deviations of the weight distributions
is indeed the major cause of the errors observed in the activity statistics,
the optimized discretization method should substantially reduce these
errors. In the $1$-bin case the standard deviation is per definition
zero and the optimization procedure cannot be applied. Similarly the
optimization procedure is not applied in the cases with $2^{16}$and
$2^{24}$ bins. Therefore, the shown data for $1,\,2^{16},\text{ and }2^{24}$
bins are the same in figures \ref{fig:indegree-naive} and \ref{fig:indegree-opt}.
Already for two bins, the FR, CV and CC distributions resulting from
the optimized discretization visually match the distributions from
the reference network in all neuronal populations in \prettyref{fig:indegree-opt}(d\textendash f)
in contrast to \prettyref{fig:indegree-naive}(d\textendash f). The
$\text{KS}$ score confirms that the optimized discretization improves
the accuracy of simulation with low numbers of bins (\prettyref{fig:indegree-opt}g\textendash i).
A discretization using two bins is sufficient to yield scores of the
same order as or even smaller than the baseline resulting from the
comparison of different realizations of the reference network. Increasing
the number of bins beyond two does not lead to any further improvements
for CV and CC. The KS score for the FR decreases slightly (not visible
here) up to around $16$ bins, from where it remains stationary for
all higher number of bins. For the \emph{fixed in-degree} version
of the PD model, the accuracy of the simulation is therefore preserved
with a $2$-bin weight discretization.

\subsection{Minimal weight resolution depends on in-degree heterogeneity}

\label{subsec:results-total-number}

\begin{figure}
\begin{centering}
\includegraphics[width=1\textwidth]{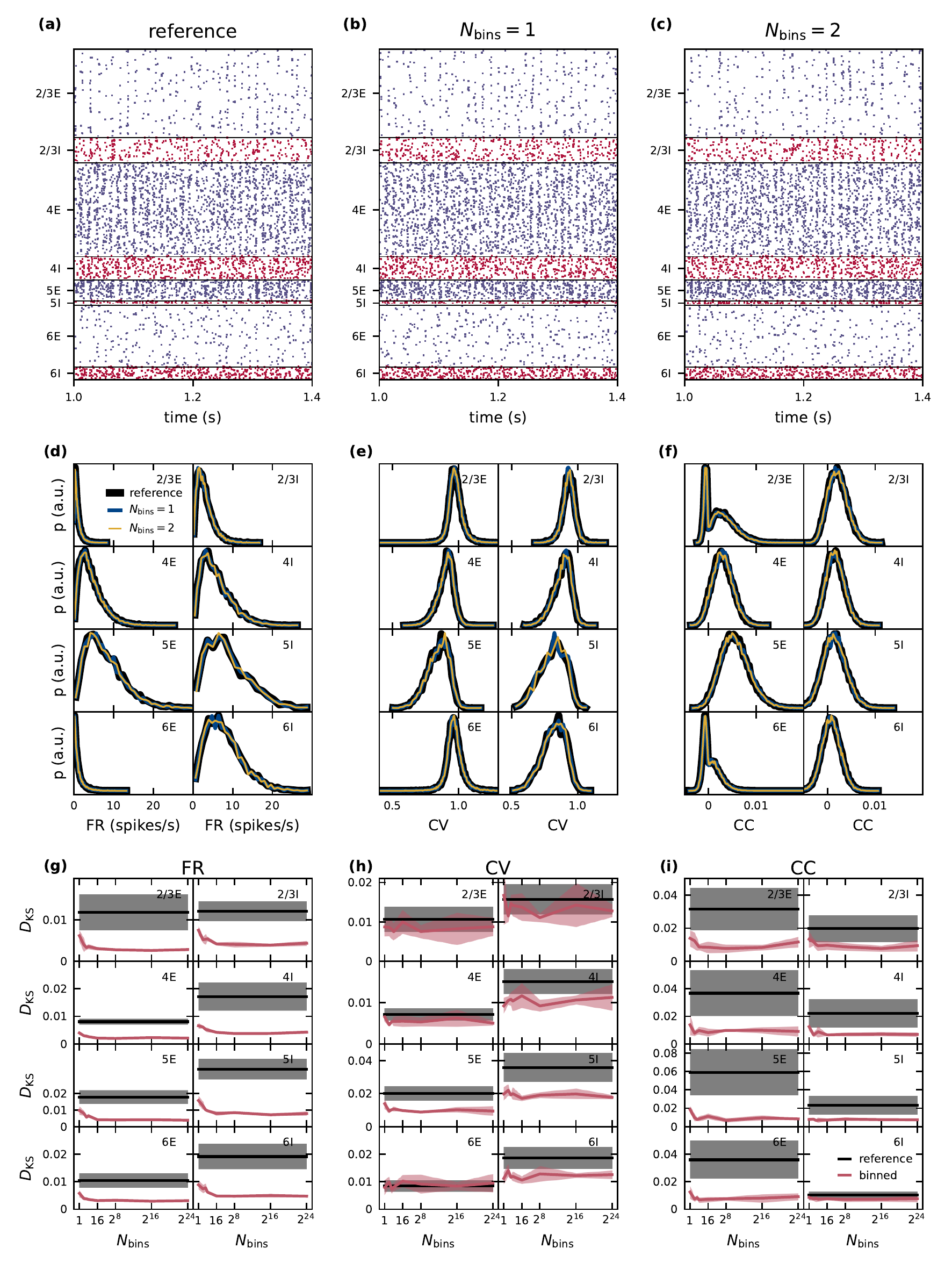}
\par\end{centering}
\caption{\textbf{Effect of optimized weight discretization on the spike-train
statistics in networks with fixed total numbers of connections. }Same
display as in \prettyref{fig:indegree-naive}. \label{fig:total-number-opt}}
\end{figure}

So far we studied the PD model with fixed in-degrees. In this section
we move on to a model version in which the neuronal populations are
connected with the \emph{fixed total number} rule (as originally used
by \citealp{Potjans14_785}), leading to binomial distributions of
the numbers of incoming connections per neuron in each population.
In comparison to the networks used in sections \ref{subsec:results-indegree}
and \ref{subsec:results-indegree-opt}, this distribution of in-degrees
leads to a heterogeneity across neurons inside one population independent
of the weight distributions. We use the optimized discretization scheme
and employ the same statistical analysis as in the previous section,
to determine how this additional network heterogeneity influences
the accuracy of network simulations subject to weight discretization.

In \prettyref{fig:total-number-opt}(d\textendash f), the distributions
of firing rates, coefficients of variation of the interspike intervals
and correlation coefficients using reference weights have different
shapes and in most populations increased widths compared to the distributions
in the previous \emph{fixed in-degree} network in \prettyref{fig:indegree-opt}(d\textendash f).
For all three statistics (FR, CV, CC) the distributions of the binned
network match those of the reference network closely for one and two
bins (\prettyref{fig:total-number-opt}d\textendash f). As before,
we quantify the deviations of the simulations with the binned weights
from the reference using the KS score (\prettyref{fig:total-number-opt}g\textendash i).
For CV and CC, the score shows no systematic trend with varying number
of bins. For the FR, there is a small descend from one to around $16$
bins and a stationary score for all higher numbers of bins. Nevertheless
for all three statistical measures the score values are always smaller
or of similar order as the disparity between different realizations
of the reference network. We conclude that already just one bin successfully
preserves the activity statistics in the PD model with distributed
in-degrees and using the optimized discretization method this remains
true also for higher numbers of bins.

\subsection{Mean-field theory relates variability of weights, in-degrees and
firing rates to minimal weight resolution}

\label{subsec:theory}

Two weight bins preserve the activity statistics of the PD model in
the \emph{fixed in-degree} network (\prettyref{subsec:results-indegree-opt})
and one weight bin is sufficient for the network with heterogeneous
in-degrees (\prettyref{subsec:results-total-number}). This observation
calls for a deeper look at the influence of weight and in-degree heterogeneity
on the firing statistics. In mean-field approximation \citep{Fourcaud02,Schuecker15_transferfunction},
the stationary firing response of a neuron $i$ with exponential postsynaptic
currents is fully determined by the first two cumulants
\begin{eqnarray}
\mu_{i} & = & \tau_{\text{s}}\sum_{j\in X_{i}}w_{ij}\nu_{j},\label{eq:mu}\\
\sigma_{i}^{2} & = & \tau_{\text{s}}\sum_{j\in X_{i}}w_{ij}^{2}\nu_{j},\label{eq:sigma}
\end{eqnarray}
of its total synaptic input current. Here, $X_{i}$ denotes the population
of neurons presynaptic to $i$, $\nu_{j}$ the stationary firing rate
of presynaptic neuron $j$, $w_{ij}$ the synaptic weight, and $\tau_{\text{s}}$
the synaptic time constant. The size of the presynaptic population
$X_{i}$ defines the in-degree $K_{i}=|X_{i}|$ of neuron $i$. Any
heterogeneity in $w_{ij}$ and $K_{i}$ (and $\nu_{j})$ leads to
a heterogeneous synaptic input statistics $\mu_{i}$ and $\sigma_{i}^{2}$,
and, in turn, to a heterogeneous firing statistics. Here, we therefore
argue that weight discretization preserves the firing statistics across
the population as long as it preserves the synaptic-input statistics
across the population. Rather than developing a full self-consistent
mathematical description of this statistics (\citealp{Roxin11_16217};
\citealp{VanVreeswijk98_1321}; \citealp{Renart10_587}; \citealp{Helias14}),
we restrict ourselves to studying the effect of weight discretization
on the ensemble statistics of the synaptic-input mean $\mu_{i}$ and
variance $\sigma_{i}^{2}$, under the assumption that the distributions
of $w_{ij}$, $K_{i}$ and $\nu_{j}$ are known. For simplicity, we
limit this discussion to the first two cumulants of the ensemble distributions,
the ensemble mean $\ensmean x$ and variance $\ensvar x$ of $\mu_{i}$
and $\sigma_{i}^{2}$ ($x\in\left\{ \mu_{i},\sigma_{i}^{2}\right\} $):
{\small{}
\begin{eqnarray}
\ensmean{\mu} & = & \tau_{\text{s}}\ensmean K\ensmean w\ensmean{\nu},\label{eq:ensmean-mu}\\
\ensmean{\sigma^{2}} & = & \tau_{\text{s}}\ensmean K\left(\ensmean w^{2}+\ensvar w\right)\ensmean{\nu},\label{eq:ensmean-sigma2}\\
\ensvar{\mu} & = & \tau_{\text{s}}^{2}\left[\ensvar K\ensmean w^{2}\ensmean{\nu}^{2}+\ensmean K\ensmean w^{2}\ensvar{\nu}+\ensmean K\ensvar w\left(\ensmean{\nu}^{2}+\ensvar{\nu}\right)\right],\label{eq:ensvar-mu}\\
\ensvar{\left(\sigma^{2}\right)} & = & \tau_{\text{s}}^{2}\left[\left(\ensvar K-\ensmean K\right)\left(\ensmean w^{2}+\ensvar w\right)^{2}\ensmean{\nu}^{2}+\ensmean K\ensmean{w^{4}}\left(\ensmean{\nu}^{2}+\ensvar{\nu}\right)\right].\label{eq:ensvar-sigma2}
\end{eqnarray}
}The above expressions rely on Wald's equation \citep{Wald44_283},
the Blackwell-Girshick equation \citep{Blackwell46_310}, and general
variance properties. Note that in previous works on heterogeneous
networks, the population variance $\ensvar{\left(\sigma^{2}\right)}$
of the input variance is often neglected \citep{Roxin11_16217,Helias14,Renart10_587}.
\citet{Roxin11_16217} moreover neglect the dependence of $\ensmean{\sigma^{2}}$
on the weight variance $\ensvar w$. While the ensemble measures in
\eqref{eq:ensmean-mu}\textendash \eqref{eq:ensvar-sigma2} can be
computed for the whole neuronal network, it is more conclusive to
use population-specific ensemble measures computed individually for
each pair of source $X$ and target population $Y$. With this approach,
$\ensmean K$ and $\ensvar K$ refer to the mean and the variance
of the number of inputs from population $X$ across all neurons in
the target population $Y$, $\ensmean w$ and $\ensvar w$ to the
mean and the variance of the weights of all connections from $X$
to $Y$, and $\ensmean{\nu}$ and $\ensvar{\nu}$ to the mean and
the variance of the firing rate across neurons in the source population
$X$. Deriving these population-specific measures is possible because
$\mu$ and $\sigma^{2}$ given in \eqref{eq:mu} and \eqref{eq:sigma},
respectively, decompose into the contributions of the different source
populations. Besides, we assume that $K_{i}$ and $w_{ij}$ are drawn
independently from their respective distributions and the rates $\nu_{j}$
are also assumed to be independent. 

For each of the ensemble measures \eqref{eq:ensmean-mu}\textendash \eqref{eq:ensvar-sigma2},
we define a discretization error 
\begin{equation}
\varepsilon_{N_{\text{bins}}}\left(x\right)=\frac{\left|x_{\infty}-x_{N_{\text{bins}}}\right|}{\left|x_{N_{\text{bins}}}\right|}\label{eq:deviation}
\end{equation}
as the normalized deviation of the measure $x_{N_{\text{bins}}}$
in a network with $N_{\text{bins}}$ weight bins from its counterpart
$x_{\infty}$ in the network with the reference weight distribution.
In \prettyref{tab:theory}, we summarize $\varepsilon_{N_{\text{bins}}}$
for all four ensemble measures to assess deviations introduced by
weight discretization to one and two bins according to the optimized
scheme. In the $2$-bin case, $\ensmean w$ and $\ensvar w$ are the
same for the reference and binned networks; in the $1$-bin case,
however, only $\ensmean w$ is preserved while $\ensvar w$ vanishes
by definition. The term $\ensmean{w^{4}}$ in \eqref{eq:ensvar-sigma2}
evaluates for the normal reference weight distribution to $\ensmean w^{4}+6\ensmean w^{2}\ensvar w+3\ensvar w^{2}$,
for one bin to $\ensmean w^{4}$, and for two bins to $\ensmean w^{4}+6\ensmean w^{2}\ensvar w+\ensvar w^{2}$.
Therefore, the results in the fourth column of \prettyref{tab:theory}
are only valid for a normal reference weight distribution, while the
first three columns are valid independent of the shape of the weight,
in-degree, or firing rate distribution. For simplicity, the rate distributions
are here assumed to be similar in the reference and the binned networks.
The mean-field theory in general relates the fluctuating synaptic
input to the distribution of output spike rates by a self-consistency
equation such that any change of parameters changes both. Here we
go with the assumption of similarity as we are interested in finding
binned networks yielding similar spiking statistics as the reference.
Consequently, $\varepsilon_{1}\left(\ensmean{\mu}\right)=\varepsilon_{2}\left(\ensmean{\mu}\right)=\varepsilon_{2}\left(\ensmean{\sigma^{2}}\right)=\varepsilon_{2}\left(\ensvar{\mu}\right)=0$,
since all these measures only depend on quantities which are the same
in networks with the reference weight distribution and their binned
counterparts. Non-zero table entries result from cases where the respective
quantities do not cancel.

\begin{table}[H]
\centering{}%
\begin{tabular}{c||c|c|c|c}
$x$ & $\ensmean{\mu}$ & $\ensmean{\sigma^{2}}$ & $\ensvar{\mu}$ & $\ensvar{\left(\sigma^{2}\right)}$\tabularnewline
\hline 
$\varepsilon_{1}\left(x\right)$ & $0$ & $j$ & $j\cdot\frac{1+f}{k+f}$ & $j\cdot\left(2+j\right)\cdot\left(1+2\frac{1+f}{k+f}\right)$\tabularnewline
\hline 
$\varepsilon_{2}\left(x\right)$ & $0$ & $0$ & $0$ & $\frac{2j^{2}}{4j+\frac{k+f}{1+f}\left(1+j\right)^{2}}$\tabularnewline
\end{tabular}\caption{\textbf{Discretization error of the synaptic-input statistics for
1- and 2-bin discretization. }Discretization error \textbf{$\varepsilon_{N_{\text{bins}}}(x)$}
as defined in \prettyref{eq:deviation} for the four ensemble measures
\eqref{eq:ensmean-mu}\textendash \eqref{eq:ensvar-sigma2} (columns),
for networks with $N_{\text{bins}}=1$ and $N_{\text{bins}}=2$ weight
bins (rows). The parameters $j:=\protect\ensvar w/\protect\ensmean w^{2}$,
$f:=\protect\ensvar{\nu}/\protect\ensmean{\nu}^{2}$ and $k:=\protect\ensvar K/\protect\ensmean K$
denote the squared variation coefficients of the synaptic weights
and the firing rates, and the Fano factor of the in-degrees, respectively.\label{tab:theory}}
\end{table}

Conventional mean-field theory captures the mean and the variance
of the input fluctuations to describe the dynamical state of a recurrent
random spiking neuronal network. This is sufficient to predict characteristics
of network dynamics like the mean spike rate, the pairwise correlation
between neurons, and the power spectrum. Therefore, if the deviations
in \prettyref{tab:theory} of $\mu$ and $\sigma^{2}$ are small,
the activity statistics in the network are expected to be preserved.
Right off the bet, the $2$-bin discretization seems more promising,
because three of the four ensemble averages considered here evaluate
to zero by definition. This holds true for any in-degree, weight or
firing rate distribution as long as the $2$-bin discretization preserves
mean and standard deviation of the weight distribution. In the $1$-bin
case, the deviation of $\ensmean{\sigma^{2}}$ still depends on the
spread of synaptic weights without any further additive terms or scaling.
In particular, the term does not depend on whether the in-degrees
are distributed or not. In networks with a large spread of synaptic
weights, a $1$-bin weight discretization is therefore always insufficient.
The third column of \prettyref{tab:theory} considers $\ensvar{\mu}$,
the variance of the means of the membrane potential across the population.
Again, the deviation of this value from the reference evaluates to
exactly zero for the $2$-bin case. For a single bin, however, a more
complex term remains. For small or no variability in the number of
incoming synapses, the deviation of $\ensvar{\mu}$ in the $1$-bin
case depends on the width of the weight distribution, but the more
the in-degrees are distributed, the smaller this dependence becomes;
for a high variability of the in-degree $k\rightarrow\infty$ with
$k:=\ensvar K/\ensmean K$ the deviation goes to zero even in the
$1$-bin case. In that case, the variability of the mean membrane
potentials caused by the distributed in-degrees is so large that the
variability of the weight distribution does not matter. The deviations
of $\ensvar{\left(\sigma^{2}\right)}$, which quantify the variability
of the magnitude of the membrane potential fluctuations across populations,
are non-zero for both $1$- and $2$-bin discretization. For one bin,
a high in-degree variability $k\rightarrow\infty$ leads to a residual
deviation $j\cdot\left(2+j\right)$ that only depends on the relative
spread $j:=\ensvar w/\ensmean w^{2}$ of the reference weight distribution.
For two bins, the respective deviation declines with an increasing
variability of the in-degrees.

\begin{figure}
\centering{}\includegraphics{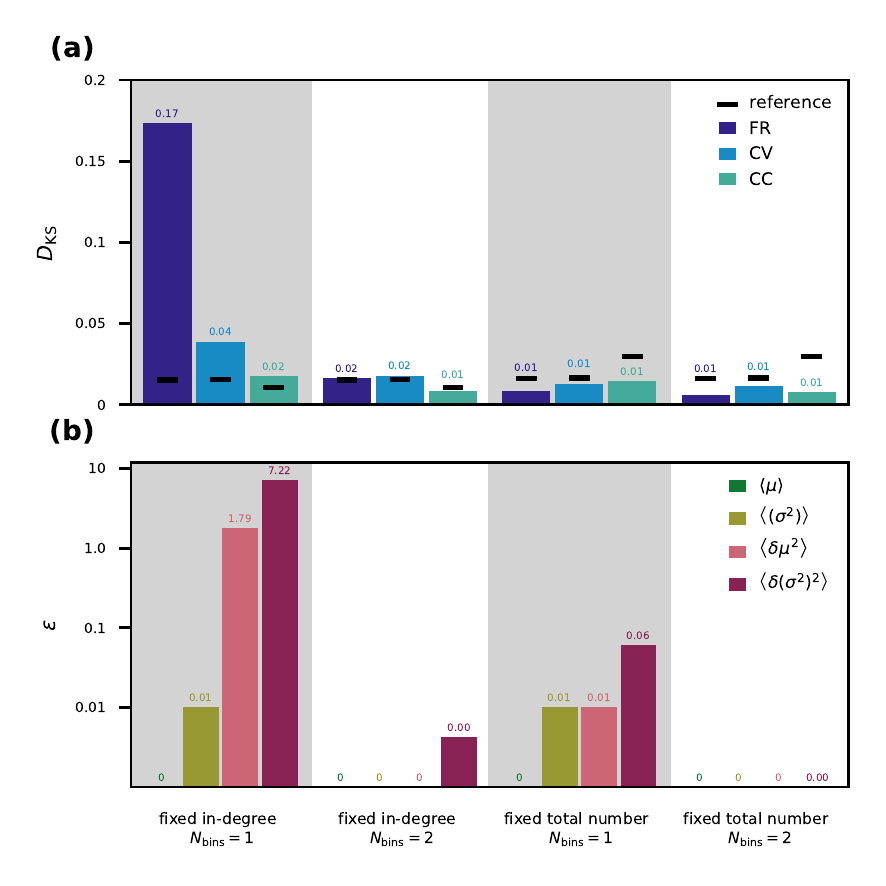}\caption{\textbf{Statistics of spiking activity and discretization errors of
the synaptic-input statistics.} (a): Mean over all neuron populations
of the $D_{\text{KS}}$ scores (for one fixed network realization)
calculated as in figures \ref{fig:indegree-opt} and \ref{fig:total-number-opt}
for FR (dark blue), CV (light blue), and CC (turquoise). Mean reference
is shown in black. (b): Mean over all pairs of source and target populations
of $\varepsilon$ values calculated as in \prettyref{tab:theory}
for the averaged input mean $\protect\ensmean{\mu}$ (green), averaged
input variance $\protect\ensmean{\sigma^{2}}$ (olive green), the
population variance of the input mean $\protect\ensvar{\mu}$ (rose),
and the population variance of the input variance $\protect\ensvar{\left(\sigma^{2}\right)}$
(purple). Logarithmic $y$-axis used for $\varepsilon$. First column:
\emph{fixed in-degree} network with $1$-bin weights. Second column:
\emph{fixed in-degree} network with $2$-bin weights. Third column:
\emph{fixed total number} network with $1$-bin weights. Fourth column:
\emph{fixed total number }network with $2$-bin weights. All $\varepsilon$
values vanishing by construction are marked as ``$0$'' without
decimals, all others are rounded to two decimal places.\label{fig:theory}}
\end{figure}

For a direct comparison of the theoretical approach and results obtained
from analyzing simulated data, we evaluate the terms in \prettyref{tab:theory}
with parameters and simulation results of our tested network models
with optimized weight discretization (figures \ref{fig:indegree-opt}
and \ref{fig:total-number-opt}). The contributions of the firing
rates are numerically computed based on measured firing rates from
simulations with the reference weight distribution. \Fref{fig:theory}
is arranged such that the KS scores of the simulated spiking activity
in panel (a) can be directly compared to the computed deviations in
the neuron input fluctuations in panel (b) for the \emph{fixed total
number} and \emph{fixed in-degree} networks with one and two weight
bins. Both the KS scores and the $\varepsilon$ values are here averaged
over populations; for completeness, \prettyref{fig:theory_suppl}
shows all $\varepsilon$ values for each pair of source and target
population individually. In the PD model \citep{Potjans14_785}, the
standard deviations of the weights are $10\%$ of the mean values,
resulting in $j=0.01$. With \emph{fixed total number} connectivity
(multapses allowed), the in-degrees are binomially distributed with
a mean of $S_{YX}/N_{Y}$ and a variance of $S_{YX}/N_{Y}\left(1-1/N_{Y}\right)$,
where $S_{YX}$ is the total number of synapses between source $X$
and target population $Y$, and $N_{Y}$ is the number of neurons
in the target population \citep{Senk20_Bernstein}. The Fano factor
of the in-degrees is therefore $k=1-1/N_{Y}$. The \emph{fixed in-degree}
scenario simplifies $k=0$ as $\ensvar K=0$.

In the \emph{fixed total number} network with one weight bin, the
normalized deviations of all of the considered ensemble measures are
very small ($\varepsilon_{1}<0.1$). This is in line with the corresponding
KS scores of the spiking activity being all below the reference. In
contrast, the \emph{fixed in-degree} network using one weight bin
exhibits large discretization errors: values above $1$ for $\ensvar{\mu}$
and even above $10$ for $\ensvar{\left(\sigma^{2}\right)}$ are observed
in some populations. These deviations explain the differences in spiking
statistics seen in \prettyref{fig:theory}(a). Using two weight bins,
all considered ensemble averages have negligible deviations corroborating
the respective observations of negligible deviations in simulation
activity statistics.

\subsection{Observation duration determines specificity of validation measures
and validation performance}

\label{subsec:sim-time}

\subsubsection{Specificity of validation measures}

\begin{figure}
\begin{centering}
\includegraphics[width=1\textwidth]{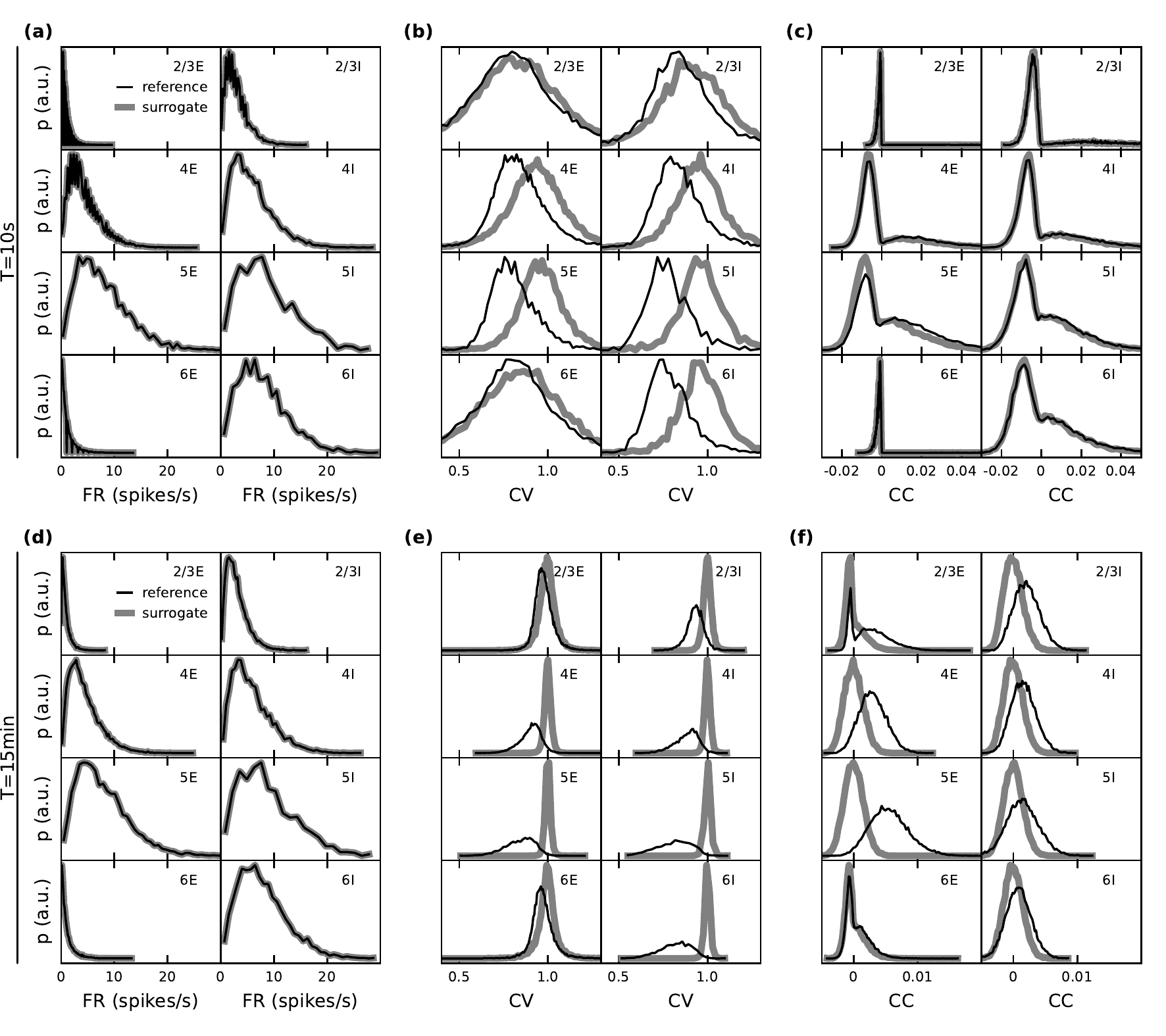}
\par\end{centering}
\caption{\textbf{Role of observation duration for the specificity of validation
measures.} Distributions of population-specific single-neuron firing
rates $\text{FR}$ (a,d), coefficients of variation $\text{CV}$ of
the interspike intervals (b,e), and spike-train correlation coefficients
$\text{CC}$ (c,f). Black: \textit{fixed total number} network model
with reference weight distribution (same as in \prettyref{fig:total-number-opt}).
Gray: surrogate data with randomized spike times (see text). Top:
observation duration $T_{\text{sim}}=\unit[10]{s}.$ Bottom: $T_{\text{sim}}=\unit[15]{min}$.
\label{fig:surrogates}}
\end{figure}

The criteria that are naturally used to validate a particular model
implementation are determined by those features the model seeks to
explain. The validation metrics should therefore reflect the specifics
of the model, rather than effects that arise from other aspects not
directly related to the model under investigation. The example of
this study, the model by \citet{Potjans14_785}, predicts that layer
and population specific patterns of firing rates, spike-train irregularities
(ISI CVs) and pairwise correlations are a consequence of the cell-type
specific connectivity within local cortical circuits. Distributions
of these quantities therefore constitute meaningful validation metrics
for this model. However, this holds only true if these distributions
are obtained such that they primarily reflect the model-specific connectivity,
and are not the result of some other trivial effects, for example
those introduced by the measurement process. A standard approach to
disentangle such effects is to compare the data generated by the model
against those generated by an appropriate null hypothesis where certain
model-specific features are purposefully destroyed (see \citet{Gruen09_1126}
for a review of methods for spiking activity and their limitations).

As an example, consider the distributions of spike-train correlation
coefficients. The PD model predicts that pairwise spike-train correlations
are small and distributed around some population-specific non-zero
mean, and that these distributions are explained by the specifics
of the connectivity. Consider now the alternative hypothesis (null
hypothesis) according to which the correlation distributions are fully
explained by the distributions of time-averaged firing rates, and
do not reflect any further characteristics of the synaptic connectivity.
An instantiation of this null hypothesis is obtained by generating
surrogate data from the model data, where the spike times for each
neuron are uniformly randomized within the observation interval. Under
this null hypothesis, the distributions of firing rates are fully
preserved (\prettyref{fig:surrogates}a and d), but the pairwise correlations
on a millisecond timescale (as well as spike-train regularities) are
destroyed. For increasing observation time $T_{\text{sim}}\to\infty$,
the distributions of correlation coefficients hence approach delta-distributions
with zero mean. For finite sample sizes, i.e., finite observation
duration $T_{\text{sim}}$, however, spurious non-zero correlations
remain. The correlation distributions obtained under this null hypothesis
therefore have some finite width, and may be hard to distinguish from
the actual model distributions. Indeed, the distributions of spike-train
correlation coefficients obtained from $T_{\text{sim}}=\unit[10]{s}$
simulations of the PD model cannot be distinguished from those generated
by the null hypothesis introduced above (\prettyref{fig:surrogates}c).
Only for sufficiently long observation intervals do the empirical
model correlation distributions carry specific information about the
network connectivity which is not already contained in the rate distributions
(\prettyref{fig:surrogates}(f) for $T_{\text{sim}}=\unit[15]{min}$).

We conclude that a model validation based on spike-train correlation
distributions should be interpreted with care: for short observation
duration (e.g., $T_{\text{sim}}=\unit[10]{s}$ as used by \citealp{VanAlbada18_291},
\citealp{Knight18_941}, \citealp{Rhodes19_20190160}, and \citealp{Golosio21_627620}),
any model implementation that preserves the rates but destroys interactions
between spike trains would not differ from the reference model with
respect to the correlation distributions. Distributions of correlations
obtained from short observation periods may however still be useful
to rule out that some model implementation erroneously generates correlations
that are significantly larger than those generated by the reference
model (see, e.g., \citealp{Pauli18}).

In principle, the same is true for other validation metrics, such
as distributions of interspike intervals (ISI) and their coefficients
of variation CVs (definitions in \prettyref{subsec:Statistics-of-spiking},
\prettyref{fig:surrogates}b and e). The surrogate data of this example
may suggest all CVs to be one, but the finite sample sizes lead to
distributions of finite widths and eventually even a shifted mean
(as seen in the $T_{\text{sim}}=\unit[10]{s}$ case). In the face
of finite observation times, one needs to check to what extent these
metrics are informative about the specifics of the underlying model,
and whether there is actually any chance that some imperfect implementation
of the model can lead to deviations from the reference. The comparison
with appropriate surrogate data is a straight forward and established
procedure to test this. In the following subsection, we employ an
alternative approach and investigate directly how the validation performance
converges with the length of the observation interval.

\subsubsection{Validation performance}

\begin{figure}
\begin{centering}
\includegraphics[width=1\textwidth]{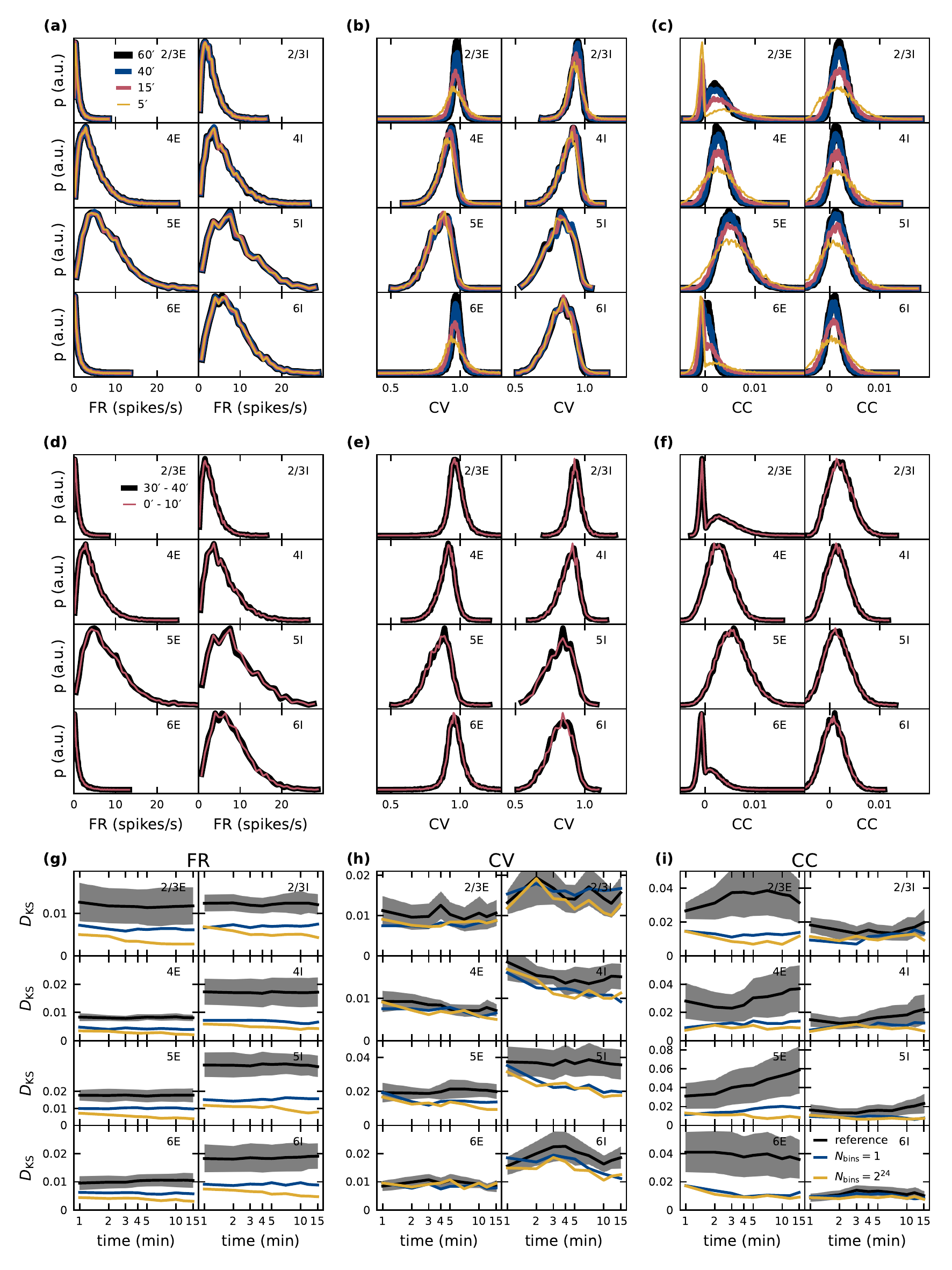}
\par\end{centering}
\caption{\textbf{Dependence of validation performance on observation interval
and duration. }Caption continued on next page. \label{fig:validation-performance-convergence}}
\end{figure}

\begin{figure}[t] \contcaption{(a)\textendash (f): Population specific
distributions of single-neuron firing rates $\text{FR}$ (a,d), coefficients
of variation $\text{CV}$ of the interspike intervals (b,e), and spike-train
correlation coefficients $\text{CC}$ (c,f) for different observation
durations ($5$, $15$, $40$, $60\,\textmd{min}$; a-c) and different
observation intervals ($[0,10]\,\textmd{min}$, $[30,40]\,\text{min}$)
with identical duration (d-f). (g)\textendash (i): Dependence of validation
performance (Kolmogorov-Smirnov score $D_{\text{KS}}$ of distributions
obtained from simulations with discretized and double-precision weights)
for single-neuron firing rates $\text{FR}$ (g), coefficients of variation
$\text{CV}$ of the interspike intervals (h), and spike-train correlation
coefficients $\text{CC}$ (i) on observation duration $T_{\text{sim}}$
with $1$-bin (blue) and $2^{24}$-bin weights (yellow). Black traces
and gray band represent mean and standard deviation of KS scores computed
with pairs of five random realizations of the reference model.} \end{figure}

Here, we tackle the question of how long simulations need to be to
yield an amount of data that is sufficient for exposing model specifics
beyond effects of finite data (see also \citet{Dahmen19_13051} for
a discussion). To investigate the convergence behavior of our validation
metrics, we analyze simulated data of up to one hour of model time
of the PD model (\prettyref{fig:validation-performance-convergence}a\textendash c).
A completely converged distribution is defined as independent of time
when its shape does not change any more if more data is added. Firing-rate
distributions converge fast; no difference is visible if analyzing
only $\unit[5]{min}$ of the data or the full hour. In contrast, the
shape of the distributions of correlation coefficients still changes
after $\unit[40]{min}$ for all populations and appears not to have
converged for the entire data recorded. The behavior of the distributions
of the coefficients of variation is population-specific: a higher
on-average firing rate leads to more spike data entering the computation
of the CVs which results in a faster convergence with simulated model
time. The convergence of distributions from low-firing neurons in
L2/3E, L2/3I and L6E, for instance, is slow.

To rule out that the underlying network dynamics change qualitatively
over time, which would have been a simple explanation for changes
in the distributions, we compare data from two $\unit[10]{min}$ intervals
separated by $\unit[20]{min}$: the distributions match for all three
metrics (\prettyref{fig:validation-performance-convergence}d\textendash f).
Convergence of the network dynamics to a stationary state happens
in fact on a much smaller time scale in the PD model network, and
we avoid distortions due to startup transients by always excluding
the very first $T_{\text{trans}}=\unit[1]{s}$ of each simulation
from the data analyzed. To achieve the same interval lengths, also
the first second of the $[30,40]\,\text{min}$ interval is excluded.

These findings make it apparent that only comparisons between simulations
of equal model time intervals are meaningful. Choosing a sufficient
length for the time intervals such that model specifics are not overshadowed
by finite-data effects is a non-trivial task which depends on the
network model itself but also on the statistical measures applied,
as shown in figures \ref{fig:surrogates} and \ref{fig:validation-performance-convergence}.
There are a couple of possible approaches for this endeavor:
\begin{enumerate}
\item The conceptually easiest way is to simulate for very long periods
of biological times (e.g., more than one hour) until all calculated
statistical distributions are converged. Because most complex neuronal
network simulations require wall-clock times much longer than the
model time simulated on modern HPC systems, this approach is unfeasible
until accelerated hardware is available \citep{Jordan18_2}.
\item Otherwise one can restrict the analysis to statistics that are less
impacted by finite data biases (e.g., the time-averaged firing rate
in \prettyref{fig:validation-performance-convergence}a). The drawback
of this approach is that a thorough validation relies on a number
of complementary metrics as decisive model-specific differences may
only become evident with some measures and not others \citep{Senk17_243}.
\item One can also derive analytical relations for the convergence behavior
of certain observables and fit them to a series of differently long
simulations. In this way the true value of the observable can be estimated
without finite data biases, as was, e.g., performed in \citet{Dahmen19_13051}.
\item The strategy employed in this study is the following: if qualitative
findings are the important parts of the study, then one can first
guess a long enough simulation time and perform the study with this.
Afterwards one has to confirm that the specific measurements to uphold
these findings are already converged also for shorter time scales
than employed in the study.  \Fref{fig:validation-performance-convergence}(g)\textendash (i)
shows the KS scores obtained for a network with \textit{fixed total
number} connections but for different simulation durations. Also for
shorter simulation times than $\unit[15]{min}$ the score of a simulation
with one bin is below the reference and therefore has acceptable accuracy,
while the improvement in accuracy when going to a high number of bins
is only small. The drawback of this approach is that one can only
confirm in retrospective if the chosen simulation time was sufficient
enough, but if one finds the opposite one would have to perform the
analysis again for longer simulation times.
\end{enumerate}

\section{Discussion}

\label{sec:discussion}

This study contributes to the understanding of the effects of discretized
synaptic weights on the dynamics of spiking neuronal networks. We
found the lowest weight resolution that can maintain the original
activity statistics for two derivatives of the cortical microcircuit
model of \citet{Potjans14_785}. In general, the discretization procedure
must preserve the moments of the reference weight distributions. In
networks where all neurons within one population receive the same
number of synaptic inputs, the variability in synaptic weights constitutes
the dominant source of input heterogeneity. In this case, the weight
discretization has to account for both the mean and the variance of
the normal reference weight distributions. In such networks, two discrete
weights are sufficient for each pair of populations to preserve the
population-level statistics. In networks where the neurons inside
the same neuronal population receive different numbers of inputs,
the variability in in-degrees may play a major role in the population-level
statistics. In the PD model with binomially distributed in-degrees,
the in-degree variability is dominating the weight variability such
that the original weights can be replaced by their mean value without
changing the population-level statistics. The study outlines a mean-field
theoretical approach to relate synaptic weight and in-degree heterogeneities
to the variability of the synaptic input statistics which, in turn,
determines the statistics of the spiking activity. We show that this
relationship qualitatively explains the effects of a reduced synaptic
weight resolution observed in direct simulations. Finally the work
sheds light on the convergence time of the activity statistics. For
a meaningful validation, the simulated model time needs to be long
enough such that the statistics are not dominated by effects of finite
sample sizes and instead are sufficiently sensitive to distinguish
model specifics from random outcomes.

In our approach synaptic weights are stored with the full floating
point resolution the computer hardware supports and all computations
are carried out using the full resolution of the floating point unit
of the processor. Discretization just refers to the fact that a synaptic
weight only assumes one of a small set of predefined values. Thus,
for the price of an indirection, only as many bits are required per
synapse as needed to uniquely identify the values in the set: one
bit for two values, 2 bit for four values. In the cortical microcircuit
model of \citet{Potjans14_785}, the recurrent weights are drawn from
one of three distinct distributions (\prettyref{subsec:network-description}).
As these weights can be replaced by the respective mean weight without
affecting the activity statistics, it is sufficient to store only
three distinct weight values (one for each synapse type) rather than
the weights for all existing synapses. Based on the $64$-bit required
for the representation of each weight of the $298,880,941$ recurrent
connections in the cortical microcircuit model with \textit{fixed
total number} connectivity, this reduces the memory demand of the
network by $\unit[2.39]{GB}$. This reduction scales linearly with
the number of synapses, such that the memory saving potential increases
for larger networks. In our reference implementation NEST this can
be achieved by using three synapses models derived from the \texttt{static\_synapse\_hom\_w}
class. If several weights are required for each group of neurons,
as is the case for \emph{fixed in-degree} connectivity, there exists
at present no pratical implementation in NEST or neuromorphic hardware
that can fully utilize a similar memory saving potential. More research
is required on suitable interfaces for the user, the domain specific
language NESTML \citep{Plotnikov16_93} offers a perspective.

The synaptic weight resolution can be substantially reduced if the
discretization procedure accounts for the statistics of the reference
weights. Simulation architectures which allow users to adapt the synaptic
weight resolution to the specific network model are therefore preferable
to those where the weight representation is fixed. This seems to advocate
the use of a mixed precision approach in neuromorphic hardware, in
which the synaptic weights are implemented with a lower resolution
while the computations are performed with higher numerical precision.
An opportunity for future development is to determine which calculation
precision is required. While it is possible to achieve comparable
network dynamics with 32-bit fixed point arithmetic \citep{VanAlbada18_291},
a minimum bit limit has not been identified, yet.

This study is restricted to non-plastic neuronal networks that fulfill
the assumptions underlying mean-field theory as presented, e.g., in
\citep{Brunel00_183}, including heterogeneous networks as studied
in \citep{Roxin11_16217}. In such networks, the distribution of synaptic
inputs across time can be approximated by a normal distribution (diffusion
approximation) such that the statistics of the spiking activity is
fully determined by the mean and the variance of this distribution.
This assumption rules out network states with low firing rates, or
correlated activity, as well as networks with strong synaptic weights.
A number of recent experimental studies revealed long-tailed, non-Gaussian
synaptic weight distributions in both hippocampus and neocortex. Here,
few individual synapses can be orders of magnitudes stronger than
the median of the weight distribution \citep[for a review, see][]{buzsaki14_264}.
Theoretical studies demonstrate that such long-tailed weight distributions
can self-organize in the presence of synaptic plasticity \citep{Teramae14_500},
and result in distinct dynamics not observed in networks of the type
studied here \citep{Teramae12_485,Iyer13_e1003248,Kriener2014_136}.
It remains to be investigated to what extent our conclusions translate
to such networks. The study by \citet{Teramae14_500} indicates that
the overall firing statistics in simple recurrent spiking neuronal
networks with long-tailed weight distributions can be preserved in
the face of a limited synaptic weight resolution, provided this resolution
does not fall below $4$ bit. Our study employs a uniform discretization
of synaptic weights with equidistant bins of identical size. For
asymmetric, long-tailed weight distributions, non-uniform discretizations
could prove beneficial. In this context, the k-means algorithm may
constitute a potential approach \citep{Muller15_arXiv}.

The connectivity of the models considered in this study is fixed
and does not change over time. If, on a given hardware architecture,
memory is scarce but computations are cheap, the connectivity of such
static networks can be implemented using an alternative approach:
connectivity data such as weights, delays, and targets do not need
to be stored and retrieved many times, but can be procedurally generated
for each spike during runtime using a deterministic pseudo-random
number generator. In particular in the case where a single synaptic
weight is sufficient to describe the projection between two populations,
the effort reduces to the procedural identification of the target
neurons. This technique has been applied, for instance, by Eugene
M. Izhikevich to simulate a large thalamocortical network model on
an HPC cluster (\url{https://www.izhikevich.org/human_brain_simulation/why.htm}),
or more recently by \citet{Knight2020_bioRxiv} to run a model of
vision-related cortical areas \citep{Schmidt18_e1006359} on GPUs,
as well as by \citet{Heittmann20_Bernstein} for a PD model simulation
using the IBM Neural Supercomputer (INC-3000) based on FPGAs. Network
models with synaptic plasticity, however, require the storage of weights
because they are updated frequently during a simulation. Plastic network
models are crucial to study slow biological processes such as learning,
brain adaptation and rehabilitation as well as brain development \citep{Morrison08_459,Tetzlaff12_715,Magee20_95}.
The present study focuses on the network dynamics at short time scales
where plasticity may be negligible. An earlier study already assessed
the effect of low weight resolutions in networks with spike-timing
dependent plasticity \citep{Pfeil12_90}. Further studies need to
investigate to what extent a reduced synaptic weight resolution compromises
the dynamics and function of plastic neuronal networks. Recent studies
indicate that good model performance could be achieved by weight discretization
methods based on stochastic rounding \citep{Gupta15_1737,Muller15_arXiv}.
Stochastic rounding could be implemented in memristive components
with probabilistic switching, thus requiring no extra random number
generators \citep{Muller15_arXiv}. It would also be interesting to
study to what extent discrete weights affect the memory capacity in
functional networks \citep{Gerstner92b,Seo11_1}. This problem is
closely linked to the question if weight discretization limits the
capabilities of neuronal networks to produce different spatiotemporal
activity patterns \citep{Kim18_e37124}. The capabilities for discretization
in functional networks depend highly on the discretization method
\citep{Senn05_61907,Gupta15_1737,Muller15_arXiv} and also the neuron
models involved. Recently, \citet{Caze20_1174} showed that non-linear
processing in dendrites enables neurons to perform computations with
significantly lower synaptic weight resolution than otherwise possible.
Therefore a principled approach to discretization methods and an adequate
selection of performance measures are necessary dependent on the respective
task.

 A large body of modeling studies treats synaptic weights as continuous
quantities that can assume any real number within certain bounds.
However, it is known since long that neurotransmission in chemical
synapses is quantized \textendash{} a consequence of the fact that
neurotransmitters are released in discrete packages from vesicles
in the presynaptic axon terminals. The analysis of spontaneous (miniature)
postsynaptic currents, i.e., postsynaptic responses to the neurotransmitter
release from single presynaptic vesicles, reveals that the resolution
of synaptic weights is indeed finite for chemical synapses. \citet{Malkin14_506},
for example, show that the amplitudes of spontaneous excitatory postsynaptic
currents recorded from different types of excitatory and inhibitory
cortical neurons are unimodally distributed with a peak at about
$\unit[20]{pA}$ and a lower bound at about $\unit[10]{pA}$. Note
that these results have been obtained despite a number of factors
that may potentially wash out the discreteness of synaptic transmission,
such as variability in vesicles sizes, variability in the position
of vesicle fusion zones, quasi-randomness in neurotransmitter diffusion
across the synaptic cleft, and variability in postsynaptic receptor
densities. For evoked synaptic responses involving neurotransmitter
release from many presynaptic vesicles, and for superpositions of
inputs from many synapses, the discreteness of synaptic strengths
is obscured and unlikely to play a particular role for the dynamics
of the neuronal network as a whole. Hence, nature, too, relies to
a large extent on discrete network connection strengths. A better
understanding of how system-level learning in nature copes with the
discrete and probabilistic nature of synapses will guide us towards
effective discretization methods for synaptic weights in neuromorphic
computers.

\section*{Conclusion}

Porting neuronal network models from multi-purpose computing systems
to neuromorphic hardware may require adjustments to the original model
description for managing hardware constraints like limited available
memory. A rigorous validation procedure assesses the effect of potential
adjustments and avoids unwanted behavior. This study makes use of
common tools from computational neuroscience including network simulation,
statistical data analysis, and a mean-field approach to challenge
relevant performance measures of a model under the assumption of a
limited synaptic weight resolution, and proposes a strategy for weight
discretization without compromising the dynamics. Future work needs
to investigate to what extent more complex networks are affected by
limiting the weight resolution. In particular, it remains an open
question whether synaptic or cell-intrinsic plasticity mechanisms
can compensate for this.

\ack{}{}

This project has received funding from the European Union\textquoteright s
Horizon 2020 Framework Programme for Research and Innovation under
Specific Grant Agreement No. 785907 (Human Brain Project SGA2) and
No. 945539 (Human Brain Project SGA3), and the Helmholtz Association
Initiative and Networking Fund under project number SO-092 (Advanced
Computing Architectures, ACA). The authors gratefully acknowledge
the computing time granted by the JARA Vergabegremium and provided
on the JARA Partition part of the supercomputer JURECA at Forschungszentrum
J\"{u}lich (computation grant JINB33).

\subsection*{Author contributions}

All authors jointly did the conceptual work, wrote the paper, reviewed
the manuscript and approved it for publication. SD performed the simulations,
analyzed and visualized the data. SD and JS developed the mean-field
theoretical approach. SD was supervised by JS and MD.

\appendix

\section{Population-specific discretization errors}

\begin{figure}
\begin{centering}
\includegraphics[width=1\textwidth]{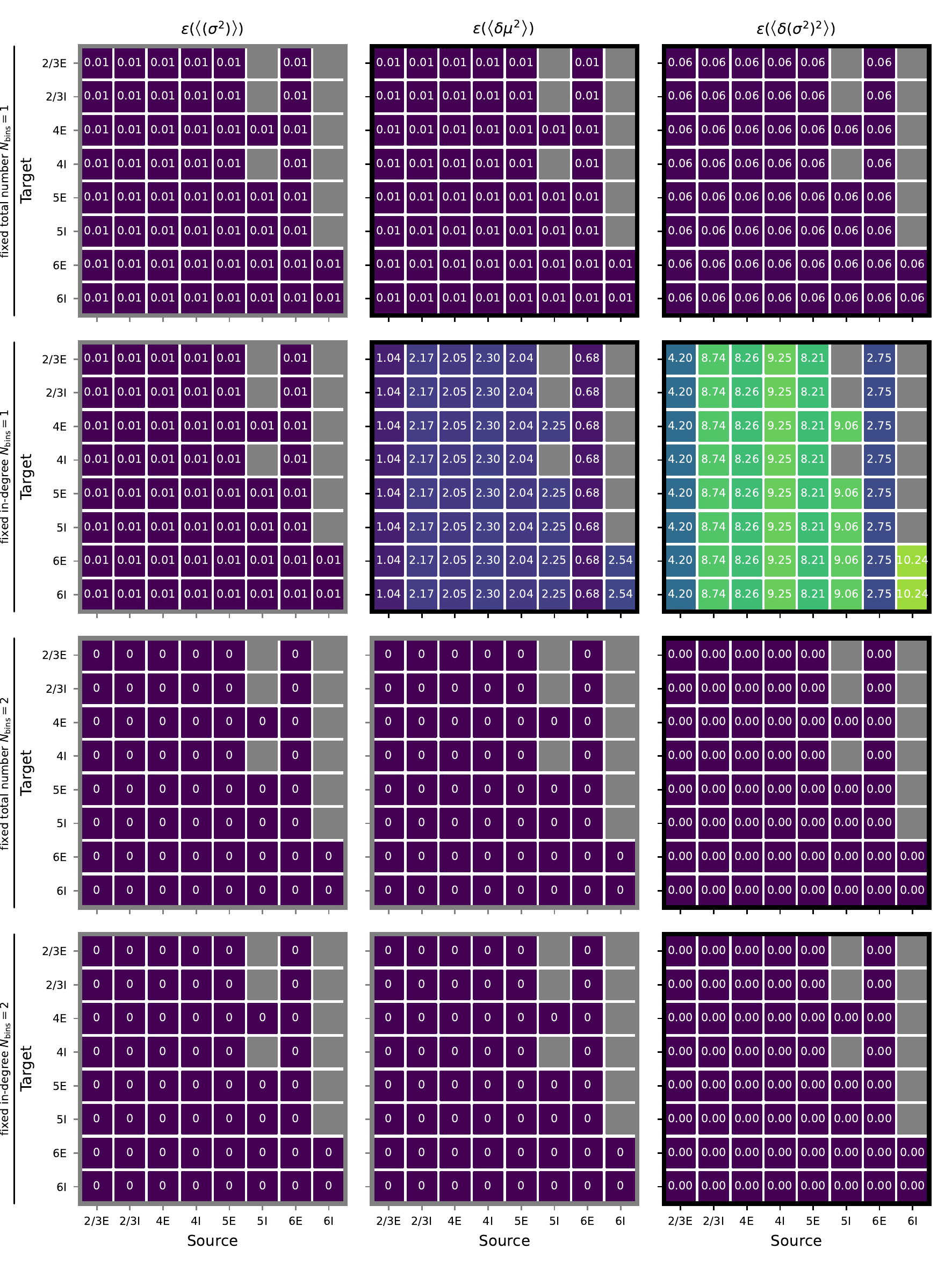}
\par\end{centering}
\caption{\textbf{Discretization errors of the synaptic-input statistics in
mean-field approximation. }Caption continued on next page. \label{fig:theory_suppl}}
\end{figure}

\begin{figure}[t] \contcaption{Discretization errors $\varepsilon$
of the population-averaged input variance $\ensmean{\sigma^{2}}$
(left column), the population variance of the input mean $\ensvar{\mu}$
(middle column), and the population variance of the input variance
$\ensvar{\left(\sigma^{2}\right)}$ (right column) for all pairs of
source and target populations. Values are calculated according to
\prettyref{tab:theory} from weight and connectivity parameters, and
from empirical firing-rate distributions obtained in network simulations.
First row: \emph{fixed total number} network with $1$-bin weights.
Second row: \emph{fixed in-degree} network with $1$-bin weights.
Third row: \emph{fixed total number} network with $2$-bin weights.
Fourth row: \emph{fixed in-degree} network with $2$-bin weights.
The discretization error of the population-averaged input mean $\ensmean{\mu}$
vanishes by construction, and is therefore not shown here. Other discretization
errors vanishing by construction are marked by ``0''. All other
errors are rounded to two decimal places. Gray matrix elements indicate
unconnected pairs of populations. Discretization errors not depending
on the firing-rate distributions are marked with gray frames.}\end{figure}

\FloatBarrier

\end{document}